\documentclass{IEEEtran}
\usepackage{cite}
\usepackage{amsmath,amssymb,amsfonts}
\usepackage{graphicx}
\usepackage{textcomp,nicefrac}
\usepackage{algorithm}
\usepackage{booktabs}
\usepackage{makecell}
\usepackage{algpseudocode}  
\usepackage[utf8]{inputenc}
\usepackage{multirow}
\usepackage{array}
\DeclareUnicodeCharacter{03B3}{\ensuremath{\gamma}}

\def\BibTeX{{\rm B\kern-.05em{\sc i\kern-.025em b}\kern-.08em
T\kern-.1667em\lower.7ex\hbox{E}\kern-.125emX}}
\begin{document}
\title{A Rapid Reconstruction Method of Gamma Radiation Field
Based on Normalized Proper Orthogonal Decomposition}

\author{Kai~Tan, Hojoon~Son, and Fan~Zhang,~\IEEEmembership{Senior Member,~IEEE}%
\thanks{All authors are with the George W.~Woodruff School of Mechanical Engineering,
Georgia Institute of Technology, Atlanta, GA 30332~USA.
(Corresponding author: Fan~Zhang, \texttt{fan@gatech.edu}).}}

\makeatletter
\def\@IEEEpubidpullup{3.4\baselineskip}
\makeatother
\IEEEpubid{\parbox[b]{\textwidth}{\centering\scriptsize
\textcopyright~2026 IEEE. Personal use of this material is permitted. Permission from IEEE must be obtained for all other uses, in any current or future media, including reprinting/republishing this material for advertising or promotional purposes, creating new collective works, for resale or redistribution to servers or lists, or reuse of any copyrighted component of this work in other works.}}

\maketitle

\begin{abstract}
In nuclear facilities, rapid and reliable reconstruction of gamma radiation fields is crucial for post-incident safety assessment and repair planning, yet direct human intervention is often impossible because of high dose rates. This task remains challenging because of unknown source distributions, complex shielding effects, and operational constraints on sensor deployment. To address these challenges, this work proposes a robot-assisted radiation field-mapping framework that equips a mobile robot with gamma detectors and couples Normalized Proper Orthogonal Decomposition (NPOD) with a minimax adaptive selection of measurement locations. By integrating the model-reduction capability of NPOD with intelligent measurement-location optimization, the proposed approach reconstructs spatially complex gamma radiation fields with high fidelity and low computational cost from sparse measurements. In a representative configuration with 160 robot stops selected from 9948 candidate spatial locations, the framework achieves a mean absolute relative reconstruction error of 0.28\%. A controlled 2-D experiment further confirms the feasibility of the method and demonstrates that the uncertainty-aware weighted least squares (WLS) formulation reduces the weighted root mean square error compared with a naive variant that ignores measurement uncertainty. Compared to the genetic algorithm baseline, the proposed adaptive measurement-location selection approach requires fewer objective evaluations and achieves a substantially lower worst-case reconstruction error without exhaustive search. Owing to its lightweight online reconstruction cost, the framework provides a practical basis for rapid radiation-field mapping and guided inspection in hazardous nuclear environments.

\end{abstract}


\section{Introduction}
\label{sec:introduction}
\IEEEPARstart{T}{he} management of radioactive materials within nuclear facilities is vital not only for operational safety but also for minimizing of the environmental and health impacts\,\cite{bangash2011,aerb2014}. Periodically monitoring the distribution of gamma rays within the facility where radiation is potentially present is one of the important aspects of radiation safety management\,\cite{iaea2011,pomaro2016,moontaha2018}. This is crucial for two reasons: 1) maintaining an updated map of radiation distribution is essential for operational safety and regulatory compliance, and 2) in the event of a radiation release or other emergencies, the radiation field can provide important information for emergency response units to make fast and effective decisions that can minimize health risks and environmental consequences. The robots equipped with gamma-radiation detectors offer clear advantages over fixed detectors. By physically navigating to positions that stationary sensors cannot cover, a robot-assisted detector system can collect data in uncontrolled or hazardous zones and adapt to the uncertainties that typically follow radiological or nuclear incidents \cite{marques2021}. Additionally, low-cost mobile radiation sensors attached to a robot can reduce the total radiation monitoring cost\cite{rahman2022,cerba2020}. However, this detection system is capable of accurately measuring radiation distributions only when they are applied at very close intervals and taking these measurements requires a large amount of measurement equipment which is time-consuming\cite{paridaens2006}. Therefore, reconstructing a gamma radiation field using sparsely spaced measurements can significantly reduce costs and improve the applicability of robot-based radiation surveying.

\IEEEpubidadjcol

Reconstructing a gamma radiation field with sparse data, particularly post-incident within a nuclear facility, presents several challenges. First, the lack of precise information about the radiation source, such as its location and intensity, impedes the use of forward modelling techniques that rely on known parameters to reconstruct the radiation field. This uncertainty often results in unreliable reconstruction outcomes. Second, the radiation map is highly nonlinear because shielding materials, such as walls or other nuclear materials, absorb and deflect gamma rays, which will complicate the prediction of radiation spread and significantly increase the complexity of the reconstruction task. Finally, the robot is limited in where it can safely navigate and collect data, resulting in sparse and potentially non-representative sampling of the radiation field. These challenges demand innovative solutions and methodologies for accurate and efficient gamma radiation field reconstruction. Currently, a variety of methods are employed to combat these challenges. Among these reconstruction methods, forward algorithms such as the Monte Carlo method\,\cite{mackin2012,kawano2010} and the point kernel method\,\cite{prokhorets2007,wang2015} simulate potential radiation scenarios based on known source configuration. Conversely, inverse algorithms, such as the interpolation and model estimation methods, are designed to deal with scenarios encountering an unknown source configuration which can most commonly be found in post-accident analyses or environments with unpredictably changing source distributions, which we address in this work. The interpolation and model estimation methods are widely applied in multiple fields. In the field of meteorology, the ensemble Kalman filter (EnKF) is utilized to enhance weather forecasting by dynamically integrating forecast models with observational data\,\cite{evensen2003}. In oceanography, singular value decomposition (SVD) is utilized to reconstruct surface temperatures and currents from satellite and buoy data, which is crucial for enhancing marine navigation and facilitating climate research\,\cite{wallace1992}. In medical imaging,  iterative reconstruction algorithms are utilized to improve the image quality by reducing noise and radiation exposure\,\cite{fessler2000}. In seismology, back-projection methods are essential for improving seismic hazard assessments by reconstructing the propagation of seismic waves\,\cite{xu2009}. In radio astronomy, the CLEAN algorithm is employed to reconstruct images of celestial bodies, allowing scientists to resolve fine details of astronomical objects\,\cite{hogbom1974}.

Interpolation techniques such as Kriging have also proven valuable in nuclear engineering to extrapolate existing data points and predict radiation levels in unmeasured areas\,\cite{sanada2015}. Zhou \emph{et al.}\ improve the interpolation process using neural networks and achieved higher accuracy in radiation pattern prediction\,\cite{zhou2021}. Khuwaileh and West \emph{et al.}\ use Gaussian process regression techniques to accurately map radiation fields with fewer measurements\,\cite{khuwaileh2020,west2021}. Zhu \emph{et al.}\ introduce a modified Cahn--Hilliard equation to reconstruct the three-dimensional gamma dose rate around a nuclear waste container, taking into account the coupling effects of multiple sources of radioactive materials and complex environmental conditions\,\cite{zhu2022ane,zhu2022pne}. Tan and Zhang use a proper orthogonal decomposition (POD) approach to reconstruct nuclear power distribution and combined it with reinforcement learning methods to determine the optimal measurement points\,\cite{tan2024}. These methods have many advantages, such as good processing capabilities in the case of sparse or incomplete data and the ability to produce high-resolution or physically consistent reconstruction results. However, they generally rely on computationally intensive and time-consuming processing. In disaster response or robot detection where rapid reconstruction is required, the development of high-performance radiation field reconstruction systems faces great challenges, mainly due to the highly nonlinear characteristics of gamma radiation field distribution.

This paper develops a novel method for reconstructing the gamma radiation field using NPOD with limited radiation measurements. In addition, a novel adaptive selection algorithm is proposed to find optimal measurement locations for rapid reconstruction. This combined method enables rapid estimation of the gamma field distribution to the point where it is feasible for a robotic system with mounted sensors to estimate the gamma distribution in real time. The developed method is also applicable to general radioactive environments. Section II introduces the detailed methodology to optimize the measurement location and rapidly reconstruct the gamma radiation fields using sparse measurements. Section III presents the results of this methodology when applied in a complex simulated radiation environment and a real experimental setting. Section IV describes the conclusion of this research.

\section{Methodology}
NPOD is utilized as a core reconstruction technique in the developed method. The NPOD method, an advanced form of the traditional POD, is specifically designed to mitigate the bias introduced by unequal variance across different measurement dimensions and enhance the fairness and accuracy of the reconstructions. In addition, an adaptive selection algorithm is proposed to optimize measurement locations to enhance the accuracy of the radiation reconstruction.
\subsection{Proper Orthogonal Decomposition}\label{sec:POD}

The primary concept underlying POD, originally formulated in fluid dynamics to analyze turbulence, aims to optimally represent the velocity field $v(x,t)\in\mathbf{H}(\Omega,\mathbf{T})$, where $\mathbf{T}\subset\mathbb{R}$; $t\in\mathbf{T}$; $x\in\Omega\subset\mathbb{R}^{n}$, $n=1,2,\ldots$. The task involves finding a deterministic function $\Phi$  in a Hilbert space $\mathbf{H}$ by resolving the maximization problem in formula (1) \cite{weiss2019}:
\begin{equation}
\max_{\Phi\in\mathbf{H}}\;
\frac{\left\langle\bigl( (v,\Phi)_{\mathbf H} \bigr)^{2}\right\rangle}
     {(\Phi,\Phi)_{\mathbf H}},
\tag{1}\label{eq:1}
\end{equation}
where $\langle\bullet\rangle$ symbolizes a statistical average operator, $(\bullet,\bullet)$ is the inner product of $\mathbf{H}$. If $\mathbf{H}$ equals to $\mathbf{L}^{2}(\Omega)$ that denotes the Hilbert space of square-integrable functions on the spatial domain $\Omega$, the maximization of problem (1) results in addressing the subsequent eigenvalue problem:
\begin{equation}
\begin{aligned}
\int_{\Omega} R(x,x')\;\Phi(x')\,\mathrm{d}x'
  &=\lambda\,\Phi(x),\\
R(x,x')
  &=\bigl\langle v(x,t)\,v(x',t) \bigr\rangle
\end{aligned}
\tag{2}\label{eq:2}
\end{equation}
where $\lambda$ is the eigenvalues. The Hilbert-Schmidt theorem assures that there exists a set of positive eigenvalues $\lambda_{1}\ge\lambda_{2}\ge\cdots\ge\lambda_{N}\ge0$ and a set of eigenmodes $\{\Phi_{1},\Phi_{2},\ldots,\Phi_{N}\}$ which is a Hilbert basis for $\mathbf{H}$. Thus, the velocity field $v$ could be decomposed according to the eigenmodes as
\begin{equation}
v(x,t)=\sum_{i=1}^{\infty}a_{i}(t)\,\Phi_{i}(x),
\tag{3}\label{eq:3}
\end{equation}
where $a_i(t)$ are the temporal coefficients, and $\bigl(\Phi_i\bigr)_{i\ge1}$ are named modes. If the given functions are discrete, such as the radiation field reconstruction using sparse measurements presented in this work, POD can be carried out by starting with constructing a matrix $\mathbf{F}\in\mathbb{R}^{M\times N}$ defined as follows,
\begin{equation}
\mathbf{F}=
\begin{bmatrix}
F_{1}(x_{1}) & F_{2}(x_{1}) & \cdots & F_{N}(x_{1})\\
F_{1}(x_{2}) & F_{2}(x_{2}) & \cdots & F_{N}(x_{2})\\
\vdots       & \vdots       & \ddots & \vdots       \\
F_{1}(x_{M}) & F_{2}(x_{M}) & \cdots & F_{N}(x_{M})
\end{bmatrix}
\tag{4}\label{eq:4}
\end{equation}
Where $\mathbf{F}$ is the snapshot matrix, $M$ is the number of spatial locations, and $N$ is the number of radiation-field snapshots in the dataset. Here, $x_i$ $(i=1,\ldots,M)$ denotes the $i$-th admissible spatial location in the radiation environment, and $\boldsymbol{x}=\{x_1,x_2,\ldots,x_M\}$ is the set of all candidate locations that the robot can navigate to within a single scenario. The corresponding correlation matrix can be defined as
\begin{equation}
\mathbf{C}=\frac{1}{M}\,\mathbf{F}^{\mathrm T}\mathbf{F}
\tag{5}\label{eq:5}
\end{equation}
The eigenpairs $(\lambda_i,\mathbf{A}_i)$ are obtained from
\begin{equation}
\mathbf{C}\,\mathbf{A}=\lambda\,\mathbf{A}
\tag{6}\label{eq:6}
\end{equation}
where $\mathbf{A}=[\mathbf{A}_{1},\mathbf{A}_{2},\ldots,\mathbf{A}_{N}]$ and $\mathbf{\Lambda}=\mathrm{diag}(\lambda_1,\ldots,\lambda_N)$. Then, the orthonormal matrix $\mathbf{\Phi}_{M\times N}$ can be calculated with Eq. (7),
\begin{equation}
\mathbf{\Phi}_{M\times N}
=\frac{1}{\sqrt{M}}\;\mathbf{F}\,\mathbf{A}\,\mathbf{\Lambda}^{-1/2}.
\tag{7}\label{eq:7}
\end{equation}
For the training snapshots, the POD coefficient of the mode $i$ in the snapshot $j$ follows from method-of-snapshots:
\begin{equation}
a_{j,i}=\sqrt{M\,\lambda_i}\;\mathbf{A}_{j,i},\qquad i=1,\ldots,N,\; j=1,\ldots,N.
\tag{8}\label{eq:8}
\end{equation}
In deployment with sparse measurements, the coefficients are inferred by least squares. Therefore, the gamma radiation field at any time $t$, which is also referred to as any condition $\textbf{F}_{t}(\textit{\textbf{x}})$, can be obtained by
\begin{equation}
\textbf{F}_{t}(\boldsymbol{x})=
\begin{bmatrix}
F_{t}(x_{1})\\ F_{t}(x_{2})\\ \vdots\\ F_{t}(x_{M})
\end{bmatrix}
=\mathbf{\Phi}_{M\times N}
\begin{bmatrix}
a_{1}(t)\\ a_{2}(t)\\ \vdots\\ a_{N}(t)
\end{bmatrix}
\tag{9}\label{eq:9}
\end{equation}
If the first $K$ orders are used, $\mathbf{F}_{t}(\boldsymbol{x})$ can also be formulated as Eq. (10):
\begin{equation}
\mathbf{F}_{t}(\boldsymbol{x})=\mathbf{\Phi}_{M\times K}
\begin{bmatrix}
a_{1}(t)\\ a_{2}(t)\\ \vdots\\ a_{K}(t)
\end{bmatrix},\quad (K<N)
\tag{10}\label{eq:10}
\end{equation}
When a POD-based reconstruction system is deployed in the real world, it has no advance knowledge of how many discrete sources are present, nor their energy and density. Therefore, to equip the POD-based reconstruction system with broad generalization, thousands of synthetic cases, such as randomly sampling the number of sources, their activities, and their positions across the domain, are generated. However, the variance of the radiation intensity grows with the square of the net source strength demonstrated analytically in the Appendix, which will cause high-activity or multi-source scenarios to possess far larger point-wise variances than those populated by weak or sparse configurations. To reduce this uncertainty, NPOD is utilized by standardizing each row of $\mathbf{F}$ by subtracting the mean and dividing by the standard deviation, thereby equalizing variance contributions from all points. Concretely, let $\mu_{i}$ and $\sigma_{i}$ be the mean and standard deviation at location $x_i$ computed from the training snapshots. Define the standardized snapshot entries and matrix by
\begin{equation}
\widetilde{F}_{j}(x_{i})=\frac{F_{j}(x_{i})-\mu_{i}}{\sigma_{i}},
\quad 1\le i\le M,\;1\le j\le N,
\tag{11}\label{eq:11}
\end{equation}
and let $\widetilde{\mathbf{F}}$ denote the matrix whose $(i,j)$ entry is $\widetilde{F}_j(x_i)$. The corresponding normalized modes $\widetilde{\mathbf{\Phi}}$ are obtained by applying Eqs.~\eqref{eq:5}--\eqref{eq:7} to $\widetilde{\mathbf{F}}$. Each location now exerts an equal influence on the decomposition due to a standardized $\widetilde{\mathbf{F}}$, which will prevent high-intensity points from monopolizing the principal modes. By balancing the variance of all input dimensions, NPOD preserves information about both strong and weak field features, significantly improving the reconstruction fidelity of low-intensity sources.  In this framework, the first step involves randomly splitting the original data into a 75\% training set and a 25\% test set. The training set is utilized to generate a NPOD model, and the measurements are normalized by mean $\mu(\mathbf{x})$ and standard deviation $\sigma(\mathbf{x})$ from the training data and then $a_{i}(t)$ is calculated. In the noise-free scenario, we can revert to the original scale by Eq. (12):
\begin{equation}
F_{t}(x_{i})=\sigma_{i}\,\widetilde{F}_{t}(x_{i})+\mu_{i}
\tag{12}\label{eq:12}
\end{equation}
However, measurement noise and a possible uniform background radiation level should be considered in real-system deployment. Since $\widetilde{F}_{t}(x_i)=\sum_{j=1}^{K}a_j(t)\widetilde{\Phi}_j(x_i)$, the measurement model on the original physical scale is
\begin{equation}
\begin{aligned}
F_t(x_i)
  &=\mu_i+\sigma_i\sum_{j=1}^{K}a_j(t)\,\widetilde{\Phi}_j(x_i)+b+\epsilon_i,\\
\mathbb{E}[\epsilon_i]
  &=0,\quad 1\le i\le M .
\end{aligned}
\tag{13}\label{eq:13}
\end{equation}
where $b $ is an unknown constant bias representing an ambient background radiation level, and $\epsilon_i$ is the measurement noise at $x_i$ which is assumed to have a zero mean and a known variance $\mathrm{Var}(\epsilon_i)$. In practice, if multiple independent readings are taken at each location $x_i$, one can use the sample mean as $F_t(x_i)$ and estimate $\mathrm{Var}(\epsilon_i)$ from the sample variance. Using multiple measurements per point in this way reduces the uncertainty of $F_t(x_i)$ because the variance of the sample mean is $\mathrm{Var}(\epsilon_i)/n_i$ where $n_i$ is the measurement times at that location.  To solve for the coefficients $a_i(t)$ and the bias $b$, a weighted least squares (WLS) approach that accounts for differing measurement uncertainties is used.

\begin{figure}[htbp]
  \centering
  \includegraphics[width=3in]{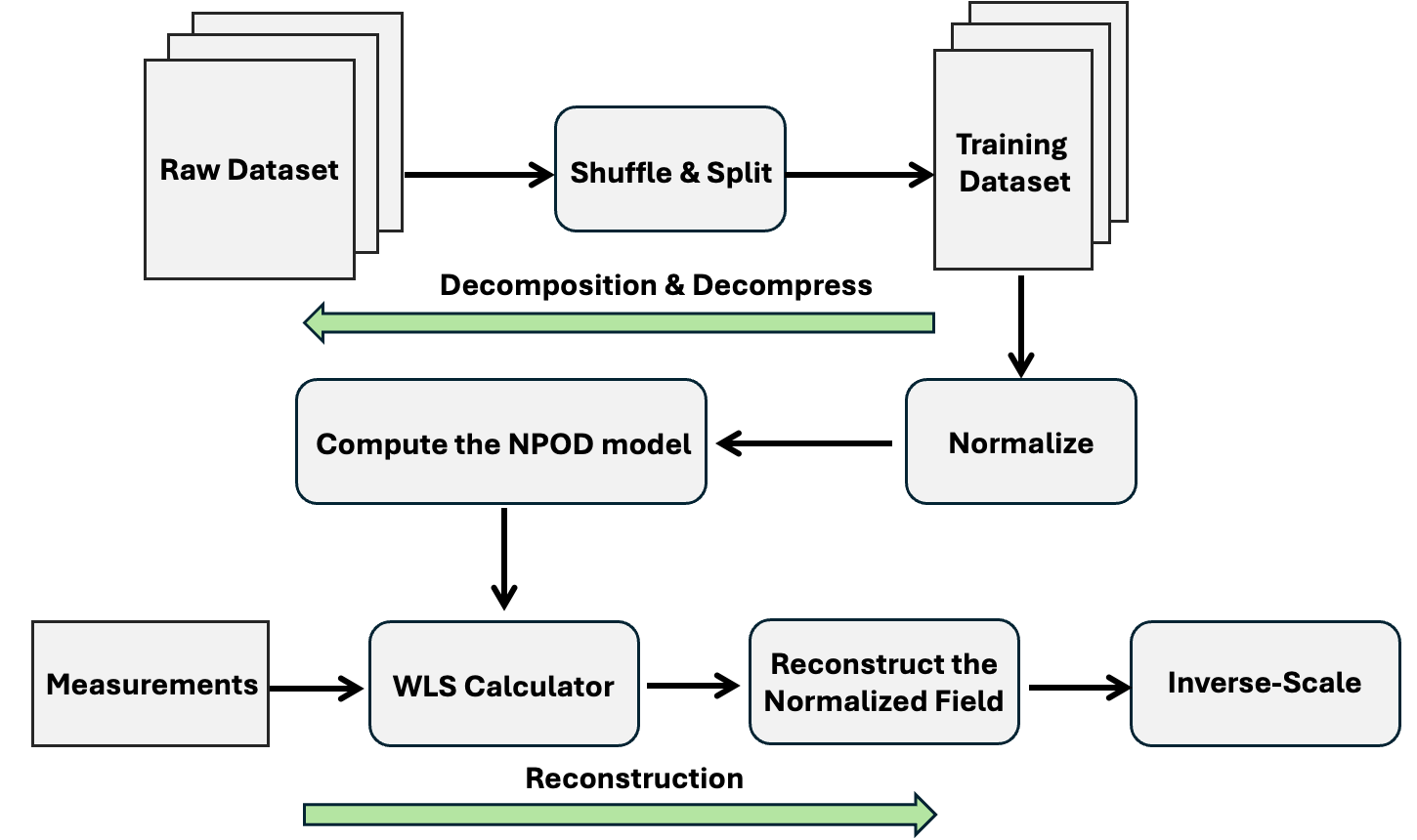}
  \caption{The flow of NPOD and WLS implementation in the reconstruction problem.}
  \label{fig:NPODflow}
\end{figure}

In WLS, each measurement equation is weighted by a factor $w_i$ that reflects the confidence of the corresponding measurement. A common and statistically motivated choice is to set the weight as the inverse of the measurement variance, $w_i=\frac{1}{\mathrm{Var}\!\left[F_t(x_i)\right]}$, so that measurements with smaller uncertainty receive larger weights and therefore exert greater influence on the solution. In practice, $F_t(x_i)$ can be obtained as the sample mean of repeated detector readings at a location $x_i$. If $\epsilon_i$ denotes the zero-mean measurement noise for a single reading and $n_i$ independent readings are collected at $x_i$, then the variance of the sample mean is $\mathrm{Var}(\epsilon_i)/n_i$, and thus $w_i$ naturally increases with repeated measurements. The WLS optimization problem is formulated by minimizing the weighted sum of squared residuals \cite{bjorck1990,fessler2000}:
\begin{equation}
\min_{\{a_j(t)\}_{j=1}^{K},\,b}\;
\sum_{i=1}^{L} w_i
\left[
F_t(x_i)-\mu_i-b
-\sigma_i\sum_{j=1}^{K}a_j(t)\,\widetilde{\Phi}_j(x_i)
\right]^2
\tag{14}\label{eq:WLS}
\end{equation}
where $L$ is the number of measurement locations. In this formulation, measurements with higher uncertainty receive smaller weights and therefore contribute less to the objective, whereas more precise measurements receive higher weights and exert greater influence on the fit. If all weights $w_i$ are equal, for example, when the measurement uncertainties are identical, or one chooses to ignore heterogeneity in noise, the WLS solution reduces to the ordinary least squares (OLS) solution. For evaluation under heterogeneous noise, a weighted root mean square error (WRMSE) is used, consistent with the WLS weighting scheme. With the original-scale prediction
$\widehat F_t(x_i)=\mu_i+\hat b+\sigma_i\sum_{j=1}^{K}\hat a_j(t)\,\widetilde{\Phi}_j(x_i)$ and measurement $F_t(x_i)$ at the $i$-th location, the WRMSE over the $L$ evaluated points is defined as
\begin{equation}
\mathrm{WRMSE}
=\left(
\frac{\sum_{i=1}^{L}
w_i\,\bigl[\widehat F_t(x_i)-F_t(x_i)\bigr]^2}
{\sum_{i=1}^{L} w_i}
\right)^{\!1/2}.
\tag{15}\label{eq:WRMSE}
\end{equation}
This definition normalizes by the total weight, so WRMSE is not affected by a uniform scaling of the weights and remains comparable across different uncertainty levels. Taking first-order optimality conditions for the WLS objective yields the normal equations \cite{bjorck1990,fessler2000}:
\begin{equation}
\mathbf{X}^T \mathbf{W} \mathbf{X}\,\hat{\mathbf{p}}
= \mathbf{X}^T \mathbf{W}\,\mathbf{y},
\tag{16}\label{eq:normal}
\end{equation}
where $\mathbf{y}=[F_t(x_1)-\mu_1,F_t(x_2)-\mu_2,\ldots,F_t(x_L)-\mu_L]^T\in\mathbb{R}^{L}$ is the mean-centered measurement vector, $\mathbf{W}=\mathrm{diag}(w_1,w_2,\ldots,w_L)$ is the diagonal weight matrix, and $\mathbf{D}_{\sigma}=\mathrm{diag}(\sigma_1,\sigma_2,\ldots,\sigma_L)$. Let $\widetilde{\mathbf{\Phi}}_L\in\mathbb{R}^{L\times K}$ denote the normalized NPOD modes restricted to the selected measurement locations, with $(\widetilde{\mathbf{\Phi}}_L)_{i,j}=\widetilde{\Phi}_j(x_i)$. The original-scale design matrix is therefore $\mathbf{X}=[\mathbf{D}_{\sigma}\widetilde{\mathbf{\Phi}}_L,\mathbf{1}_L]\in\mathbb{R}^{L\times(K+1)}$, where $\mathbf{1}_L$ is the column corresponding to the constant physical-scale bias $b$. The parameter vector is $\hat{\mathbf{p}}=
[\hat{a}_1(t),\hat{a}_2(t),\ldots,\hat{a}_K(t),\hat{b}]^T$, where the first $K$ entries are the reconstructed NPOD coefficients and the last entry is the estimated background bias. Provided that $L>K$ and the chosen measurement locations make $\mathbf{X}$ full rank, the matrix $\mathbf{X}^T \mathbf{W} \mathbf{X}$ is invertible. In that case, the solution to Eq.~(16) is obtained in the form as
\begin{equation}
\hat{\mathbf{p}} = \left(\mathbf{X}^T \mathbf{W} \mathbf{X}\right)^{-1}\mathbf{X}^T \mathbf{W}\mathbf{y}.
\tag{17}\label{eq:17}
\end{equation}
The estimate $\hat{\mathbf{p}}$ provides the inferred coefficients $\hat{a}_1(t),\ldots,\hat{a}_K(t)$ for the NPOD modes, as well as the estimated constant offset $\hat{b}$. Incorporating the constant term $b$ ensures that any uniform ambient background radiation level is accounted for in the reconstruction. If an independent background measurement is available or the background is known to be negligible, one may omit the bias term to reduce the number of fitted parameters. Otherwise, including $b$ improves flexibility and typically yields more accurate reconstructions under realistic field conditions.  Fig.~\ref{fig:NPODflow} summarizes the process of using the NPOD method for reconstructing a gamma radiation map. Finally, the number of modes can be determined by the cumulative energy ratio, defined as\cite{chen2013}
\begin{equation}
E_{K} = \frac{\sum_{i=1}^{K} \lambda_i}{\sum_{i=1}^{K_{\max}} \lambda_i},
\tag{18}\label{eq:18}
\end{equation}
Where $K_{\max} \leq \min(M, N)$ is the maximum retained rank.  $K$ is usually selected such that $E_{K}$ meets a desired threshold and ensures an appropriate balance between model complexity and accuracy.

\subsection{Minimax Adaptive Selection of Measurement Locations for NPOD-Based Reconstruction}\label{sec:adaptive}

The combination of measurement locations has a significant impact on the performance of NPOD-based reconstruction systems. At the same time, environmental complexity, robot mobility limitations, and radiation-dose constraints limit the number and placement of samples. Selecting optimized locations is therefore crucial for achieving high reconstruction accuracy with few measurements. For this combinatorial problem, advanced metaheuristics such as genetic algorithms, particle swarm optimization, and simulated annealing can iteratively optimize sampling locations. These methods account for interactions between environmental features and radiation propagation, but the search space in real scenarios is often so large that satisfactory solutions cannot be found within limited computational resources. For example, an environment with 10000 candidate grid points and a budget of 100 measurements already induces a vast number of combinations.

To mitigate this difficulty, we pose selection as a minimax problem guided by a reverse-engineering observation that OLS minimizes discrepancy only at sampled locations and may fail to guarantee global accuracy in complex fields\,\cite{bjorck1990}. In addition, the safety-critical mapping penalizes worst-case underestimation. Minimizing the maximum absolute relative error directly controls the largest relative miss over the domain, which better safeguards hotspots and thin high-gradient corridors than a mean objective. At iteration $n$, the NPOD coefficients on the current measurement location set $\mathcal{S}_n$ are obtained by
\begin{equation}
\mathbf{a}^{(n)}
=\arg\min_{\mathbf{a}\in\mathbb{R}^{K}}
\sum_{x\in\mathcal{S}_{n}}
\Bigl[F(x)-\sum_{\ell=1}^{K}a_{\ell}\,\varphi_{\ell}(x)\Bigr]^{2},
\tag{19}\label{eq:LSM}
\end{equation}
and the relative residual and NPOD field are
\begin{equation}
e_{n}(x)=\frac{F(x)-F^{(n)}(x)}{F(x)},
\qquad
F^{(n)}(x)=\sum_{\ell=1}^{K}a_{\ell}^{(n)}\varphi_{\ell}(x).
\tag{20}\label{eq:residual}
\end{equation}
The worst location and its value are
\begin{equation}
x_{n}^{\star}=\arg\max_{x\in\Omega}\bigl|e_{n}(x)\bigr|,
\qquad
\mathrm{MaxE}^{(n)}=\bigl|e_{n}(x_{n}^{\star})\bigr|.
\tag{21}\label{eq:maxErr}
\end{equation}
The objective is minimizing
\begin{equation}
J(\mathcal{S})=\max_{x\in\Omega}\,\bigl|e_{\mathcal{S}}(x)\bigr|.
\tag{22}\label{eq:minimax}
\end{equation}
In our workflow, the NPOD rank $K$ is first selected offline from the snapshot ensemble using the cumulative energy criterion, and Algorithm~\ref{alg:adaptive-acc} is then applied to optimize the measurement locations for that fixed $K$. Numerical stability is ensured by solving  $X^\top X+\alpha I$ for a small ridge parameter $\alpha\ge 0$, combined with stable Cholesky factorization, where it recovers OLS when $\alpha=0$. The procedure integrates a greedy growth phase with a restricted single-swap refinement, summarized in Algorithm~\ref{alg:adaptive-acc}. From an initial set $\mathcal{S}_0$ with $|\mathcal{S}_0|\ge K$ constructed by variance ranking, leverage ranking, or random sampling without replacement, we apply a lightweight ranking to unsampled candidates. A convex combination
\[
\mathrm{score}(x)=w_{\mathrm{err}}\,\widetilde{E}(x)+w_{\mathrm{lev}}\,\widetilde{\Lambda}(x)+w_{\mathrm{var}}\,\widetilde{V}(x),
\tag{23}\label{eq:score}
\]
is used, where $\widetilde{E}(x)$ is the current row-wise maximal relative error normalized to $[0,1]$, $\widetilde{\Lambda}(x)$ is the leverage of the NPOD basis at location $x$, and $\widetilde{V}(x)$ is the empirical variance at $x$. The nonnegative weights $w_{\mathrm{err}},w_{\mathrm{lev}},w_{\mathrm{var}}$ satisfy $w_{\mathrm{err}}+w_{\mathrm{lev}}+w_{\mathrm{var}}=1$. The top $P$ points by $\mathrm{score}(\cdot)$ define a candidate pool $\mathcal{C}\subset\Omega\setminus\mathcal{S}_n$. For each $c\in\mathcal{C}$, the exact objective $J(\mathcal{S}_n\cup\{c\})$ is evaluated by reusing factorizations and computing updates in batches of size $B$ to control memory and exploit efficient linear-algebra kernels. The best candidate $c^\star$ is appended to $\mathcal{S}_n$. Ties are resolved by the mean relative error, and a larger smallest singular value $\sigma_{\min}$ of the POD submatrix is selected when still tied. The greedy loop continues until the point budget $N_p$ is reached or a target error is achieved. An early-abort tolerance short-circuits evaluation of a candidate once it becomes provably worse than the current best. After obtaining a full set $\mathcal{S}$, a restricted single-swap refinement is applied. Let $I$ denote the number of in-set indices examined, and $O$ the size of the out-of-set pool ranked by $\mathrm{score}(\cdot)$. For each pair $(s,c)$ with $s$ from these $I$ indices and $c$ from these $O$ candidates, we compute $J\bigl((\mathcal{S}\setminus\{s\})\cup\{c\}\bigr)$ in batches and accept the best improving swap. Iterations stop when no swap decreases $J$ by more than a small threshold $\delta$ or when a maximum number of swap rounds is reached. Robustness is enhanced through repeating the full pipeline from several initializations and returning the set that minimizes $J(\mathcal{S})$. Near ties are resolved by the mean relative error and $\sigma_{\min}$ is reported as a stability proxy.

\begin{algorithm}[htbp]
  \caption{Minimax adaptive selection of measurement locations (candidate pool, batching, and restricted single-swap)}
  \label{alg:adaptive-acc}
  \begin{algorithmic}[1]
    \State \textbf{Input:} domain $\Omega$, POD rank $K$, budget $N_p$, targets $\mathrm{MaxE}_{\mathrm{T}}$ and $\delta$, ridge $\alpha\ge 0$, candidate-pool size $P$, batch size $B$, swap budgets $I$ and $O$, restarts $R$, and weights $w_{\mathrm{err}},w_{\mathrm{lev}},w_{\mathrm{var}}$ with $w_{\mathrm{err}}+w_{\mathrm{lev}}+w_{\mathrm{var}}=1$.
    \For{$r=1$ to $R$}
        \State Initialize $\mathcal{S}_{0}$ with $|\mathcal{S}_{0}|\ge K$ using variance ranking, leverage ranking, or random sampling.
        \State \emph{Greedy growth with candidate pool (exact $J$ via batched updates)}
        \While{$|\mathcal{S}_{n}|<N_p$}
            \State Rank $\Omega\setminus\mathcal{S}_{n}$ by $\mathrm{score}(\cdot)$ and form $\mathcal{C}$ with $|\mathcal{C}|=P$.
            \State Evaluate $J(\mathcal{S}_{n}\cup\{c\})$ for $c\in\mathcal{C}$ in batches of size $B$ (with early-abort).
            \State Append $c^\star=\arg\min_{c\in\mathcal{C}} J(\mathcal{S}_{n}\cup\{c\})$; update $n$.
            \If{$\mathrm{MaxE}^{(n)}\le\mathrm{MaxE}_{\mathrm{T}}$ \textbf{or} $\bigl|\mathrm{MaxE}^{(n)}-\mathrm{MaxE}^{(n-1)}\bigr|\le\delta$}
                \State \textbf{break}
            \EndIf
        \EndWhile
        \State \emph{Restricted single-swap refinement}
        \Repeat
            \State Select $I$ in-set indices with largest row-wise errors and $O$ out-of-set indices by $\mathrm{score}(\cdot)$.
            \State Search $(s,c)$ pairs within these lists in batches; accept the swap with the largest drop in $J$.
        \Until{no swap reduces $J$ by more than $\delta$ or a maximum swap count is reached}
        \State Store $\mathcal{S}$ and $J(\mathcal{S})$.
    \EndFor
    \State \textbf{Output:} $\mathcal{S}_{\ast}$ minimizing $J(\mathcal{S})$ across runs; ties resolved by mean relative error; report $\sigma_{\min}$.
  \end{algorithmic}
\end{algorithm}

\subsection{Genetic Algorithms}\label{sec:GA}

\begin{algorithm}[!b]
  \caption{Genetic Algorithms}
  \label{alg:GA}
  \footnotesize
  \begin{algorithmic}[1]
    \State \textbf{Input:} $N_{pop}$, $\mathrm{MuR}$, $\textit{max\_gen}$, NPOD model with certain number of modes, $\mathrm{Ep}$
    \State Initialize population with $N_{pop}$ individuals, each individual is a binary vector representing positions.
    \State Initialize generation counter $g = 1$
    \While{$g \le \textit{max\_gen}$}
        \State Compute fitness of each individual in the population based on the reconstruction results with NPOD.
        \State Sort individuals based on fitness and select $\mathrm{Ep}$ of the population as elites.
        \State Perform roulette wheel selection to fill the new generation.
        \State Apply crossover and mutation operations to create new individuals until the population reaches $N_{pop}$.
        \State Replace the least fit individuals with elites to maintain $\mathrm{Ep}$.
        \State Update population with new generation.
        \State $g = g + 1$
    \EndWhile
    \State Select the best individual based on fitness after all generations.
    \State Decode the best individual to determine the optimal measurement positions.
    \State \textbf{Return} the best individual (i.e.\ the optimal measurement positions).
  \end{algorithmic}
\end{algorithm}

Genetic Algorithm (GA) draws on the natural selection and inheritance mechanism of the biological world and can efficiently explore and find approximate optimal solutions in complex high-dimensional search spaces\,\cite{holland1992}. Due to its good adaptability to nonlinear constraint problems, GA is often used as a benchmark for evaluating new optimization methods. Algorithm~\ref{alg:GA} shows the basic process of GA for screening the optimal measurement point configuration. First, a set of candidate solutions is randomly generated as the initial population, and each candidate solution can be regarded as a chromosome containing several ``genes''. Then, in each generation, the population performance is evaluated through the fitness function, and excellent individuals are selected for crossover and mutation to reproduce the next generation. The performance of the algorithm is mainly controlled by the following hyperparameters. The population size $N_{\mathrm{pop}}$ determines the number of candidate solutions for parallel search. The larger the population size, the richer the genetic diversity, which helps to escape the local optimum. The mutation rate, $\mathrm{MuR}$, randomly changes some genes in the solutions, ensuring that the algorithm does not get stuck in local optima. Meanwhile, the maximum number of generations, $\textit{max\_gen}$, controls when the GA stops, striking a balance between thorough exploration and computational feasibility. The elite percentage, $\mathrm{Ep}$, helps preserve quality solutions by passing a fixed percentage of top-performing individuals unchanged into the next generation.

\section{Results}\label{sec:results}

\begin{figure}[htbp]
  \centering
  \includegraphics[width=2.6in]{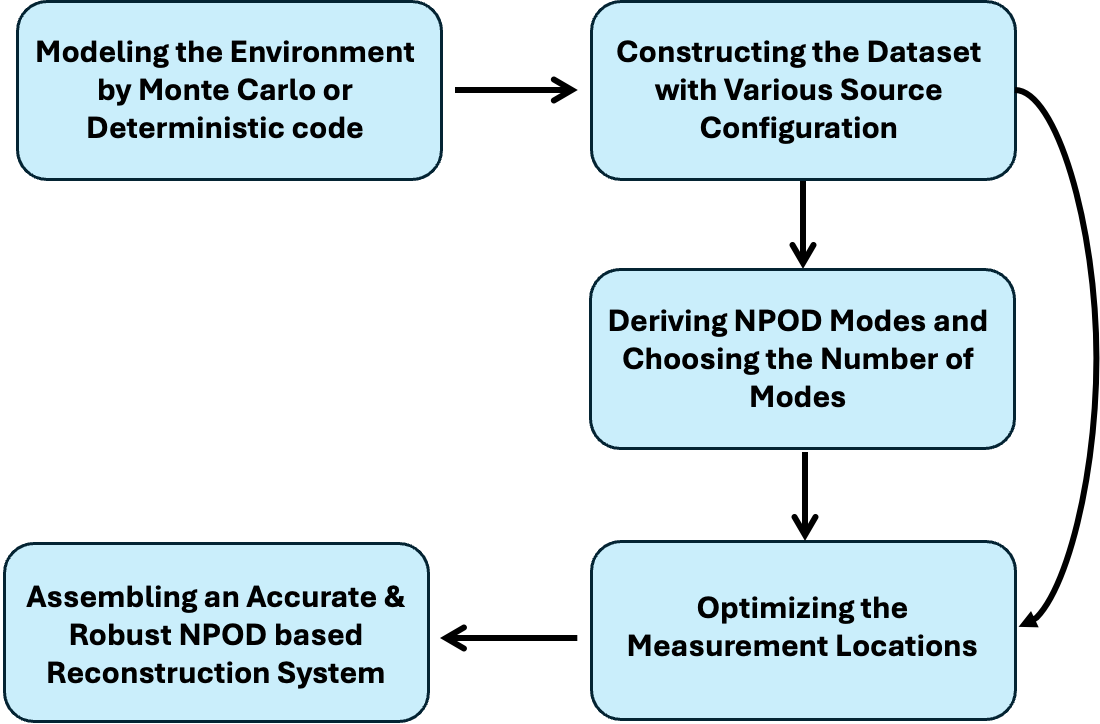}
  \caption{NPOD-based gamma-radiation-field reconstruction framework.}
  \label{fig:framework}
\end{figure}

In this study, a framework for gamma radiation field reconstruction based on the NPOD is developed, as shown in Fig.~\ref{fig:framework}. This framework comprises two primary components: 1) the construction of
the NPOD model, and 2) the determination of an optimal configuration for measurement locations based on the location selection algorithm. In terms of robustness, the dataset is expanded to encompass a wide variety of  gamma radiation distributions, which include multiple gamma sources in multiple locations. This approach ensures that the NPOD-based framework is equipped to deliver high performance across diverse scenarios, including fields generated by various numbers and configurations of radiation sources. To validate the effectiveness and resilience of the proposed methods, a large-scale simulation case including both noise-free and experiment-calibrated noisy reconstructions, and a controlled 2-D laboratory experiment are utilized.  In the verification setup, the NPOD basis is precomputed offline from synthetic snapshots and kept fixed during deployment. Onboard inference solves a small WLS problem for $(a_{1}, \ldots, a_{K}, b)$ using the $L$ measurements. This process does not require retraining the NPOD basis for each scene.  Notably, the measurements can be provided either by a mobile robot visiting selected locations or by a fixed detector network, since the online reconstruction only requires the set of $L$ readings and the precomputed NPOD basis. In practice, a robot is convenient when many locations must be sampled, while fixed detectors are useful for a small set of critical or hard-to-reach points. A hybrid setup that combines both can improve coverage under mobility constraints.

\subsection{Case 1: Simulation-Based Verification}

\begin{figure}[htbp]
  \centering
  \includegraphics[width=3in]{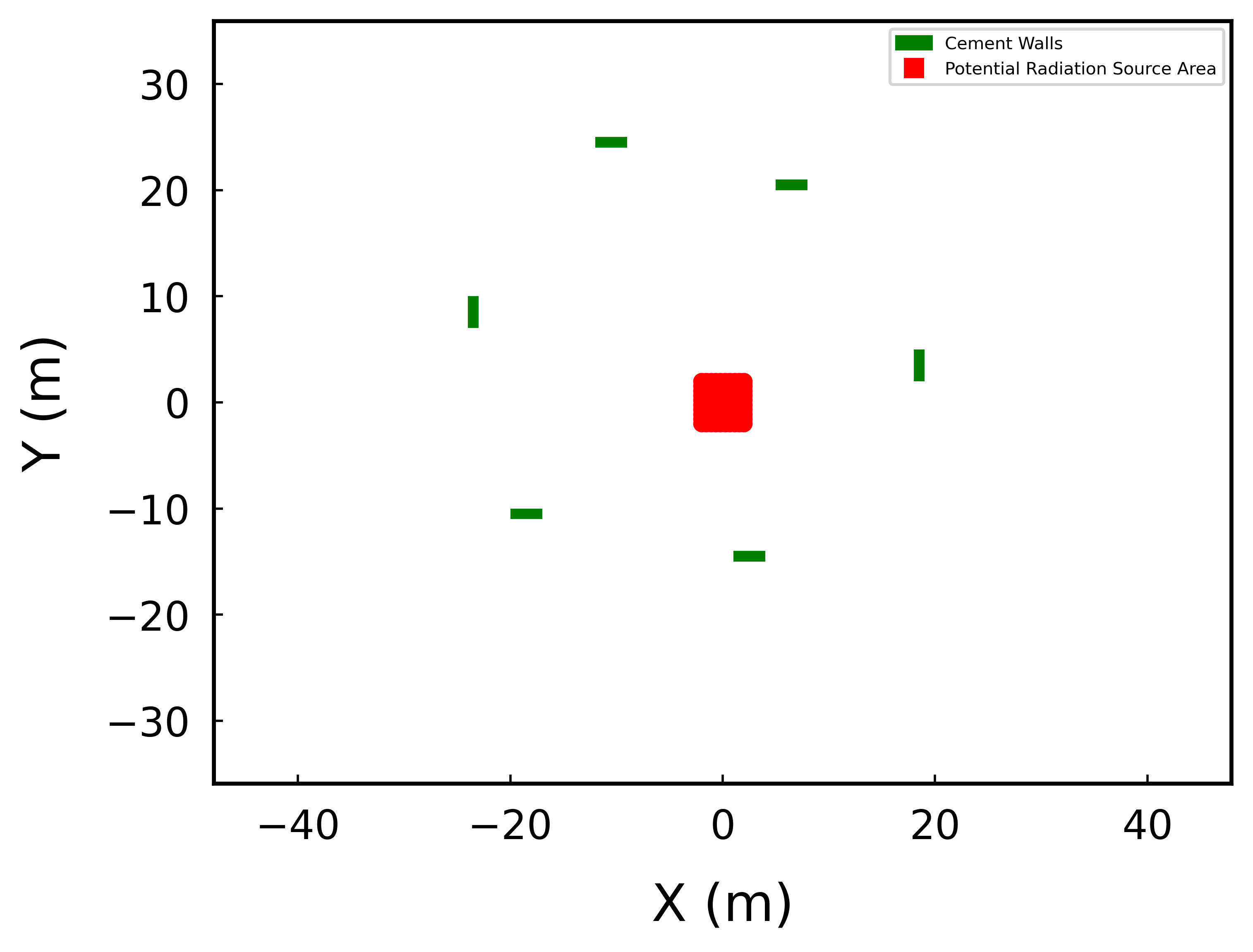}
  \caption{The layout of the test case}
  \label{fig:geometry}
\end{figure}

This scenario involves a dense array of obstacles and a complex configuration of radiation sources, leading to a highly intricate gamma-radiation distribution. It is used to verify the ability of the proposed framework to capture such complex patterns and to demonstrate that the measurement selection algorithm can effectively handle a larger search space. The scenario's testing ground covers an area of $96~\mathrm{m} \times 72~\mathrm{m}$, as detailed in Fig.~\ref{fig:geometry} and Table~\ref{tab:geom}. It mimics a setting with air-filled spaces and concrete walls, representing common materials found in nuclear facilities.  To enhance the robustness of the framework, a strategy for selecting various configurations of radiation sources is employed. It involves first deciding on the number of sources, followed by placing these sources randomly within a designated source area highlighted in red. The configurations with 1, 2, 4, 6, 10, 15, 20, 40, 60, and 100 sources and 450 cases for each are examined. In total, $N=4500$ distinct scenarios are created for analysis, each with a specific number of sources. If prior information of the source is available, such as expected source counts or likely hot zones, the sampling is biased accordingly while maintaining a low-probability tail to ensure generalization.  The snapshot generator explicitly spans numerous source multiplicities and placements to ensure the NPOD basis capture patterns arising from single to dense multi-source fields.  The simulation of gamma radiation flux across these scenarios is conducted using OpenMC\,\cite{romano2015}, with the vacuum boundary simulation condition and a spatial resolution of 100~$\times$~100 mesh points for each scenario. After excluding the source region and the grid cells occupied by obstacles, the remaining admissible lattice points that the sensor platform can access are $M=9948$. This resolution is selected to adequately resolve the dominant geometric features in the environment. For example, the wall and source-region dimensions are on the meter scale. Therefore, a $100 \times 100$ mesh provides sub-meter cell sizes over the $96~\mathrm{m} \times 72~\mathrm{m}$ area and captures the relevant spatial gradients. In addition, a mesh-independence verification is performed to confirm that further refinement does not materially change the tallied field.

\begin{table}[!t]
\caption{Parameters of the geometry used in the test case shown in Fig.~\ref{fig:geometry}.}
\label{tab:geom}
\centering
\small
\renewcommand{\arraystretch}{1.15}
\begin{tabular}{@{}lccc@{}}
\toprule
Object        & Location $(x,y)$  & Height (m) & Width (m) \\
\midrule
Wall\,1       & $(-23.5,\;8.5)$   & 3.0 & 1.0 \\
Wall\,2       & $(-18.5,-10.5)$   & 1.0 & 3.0 \\
Wall\,3       & $(2.5,-14.5)$     & 1.0 & 3.0 \\
Wall\,4       & $(18.5,\;3.5)$    & 3.0 & 1.0 \\
Wall\,5       & $(6.5,\;20.5)$    & 1.0 & 3.0 \\
Wall\,6       & $(-10.5,\;24.5)$  & 1.0 & 3.0 \\
Source region & $(0,\;0)$         & 4.0 & 4.0 \\
\bottomrule
\end{tabular}
\end{table}

\begin{figure}[htbp]
  \centering
  \includegraphics[width=3in]{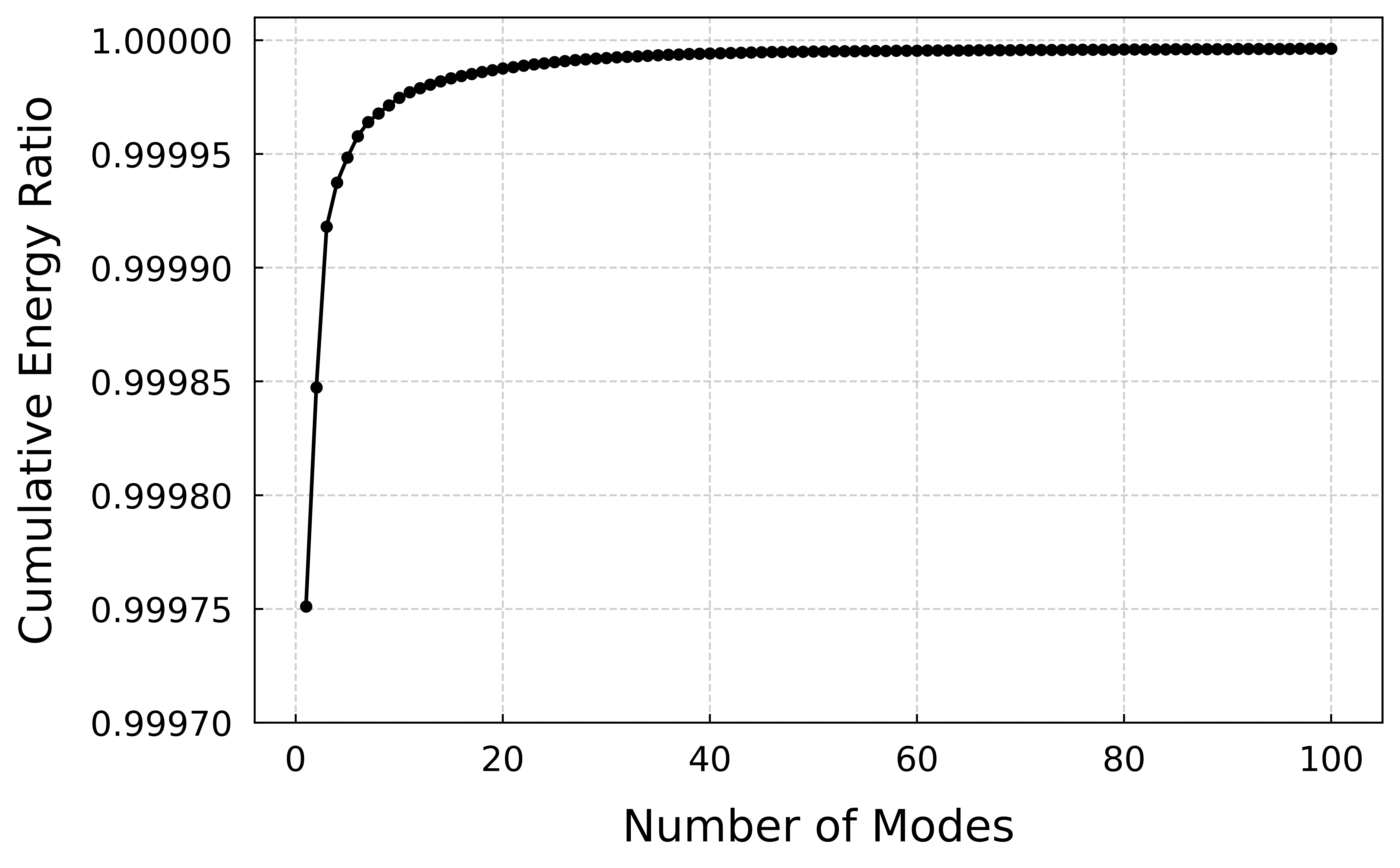}

  \caption{The cumulative energy distribution of NPOD modes.}
  \label{fig:cumEnergy}
\end{figure}

Fig.~\ref{fig:cumEnergy} presents the cumulative energy distribution of NPOD modes, showing how the cumulative energy ratio $E_K$ evolves as the number of modes increases. The value of  $E_K$ approaches 1, indicating nearly complete representation of the system's energy, when the number of modes reaches about 70. Beyond this point, additional modes produce only marginal improvements, and in practice $E_K > 99.9\%$ is sufficient~\cite{macraild2024}. Therefore, 70 is chosen as the optimal number of modes for the NPOD.

\subsection{Optimization of Measurement Location and Reconstruction Results}

\begin{figure}[!t]
\centering
\includegraphics[width=\columnwidth]{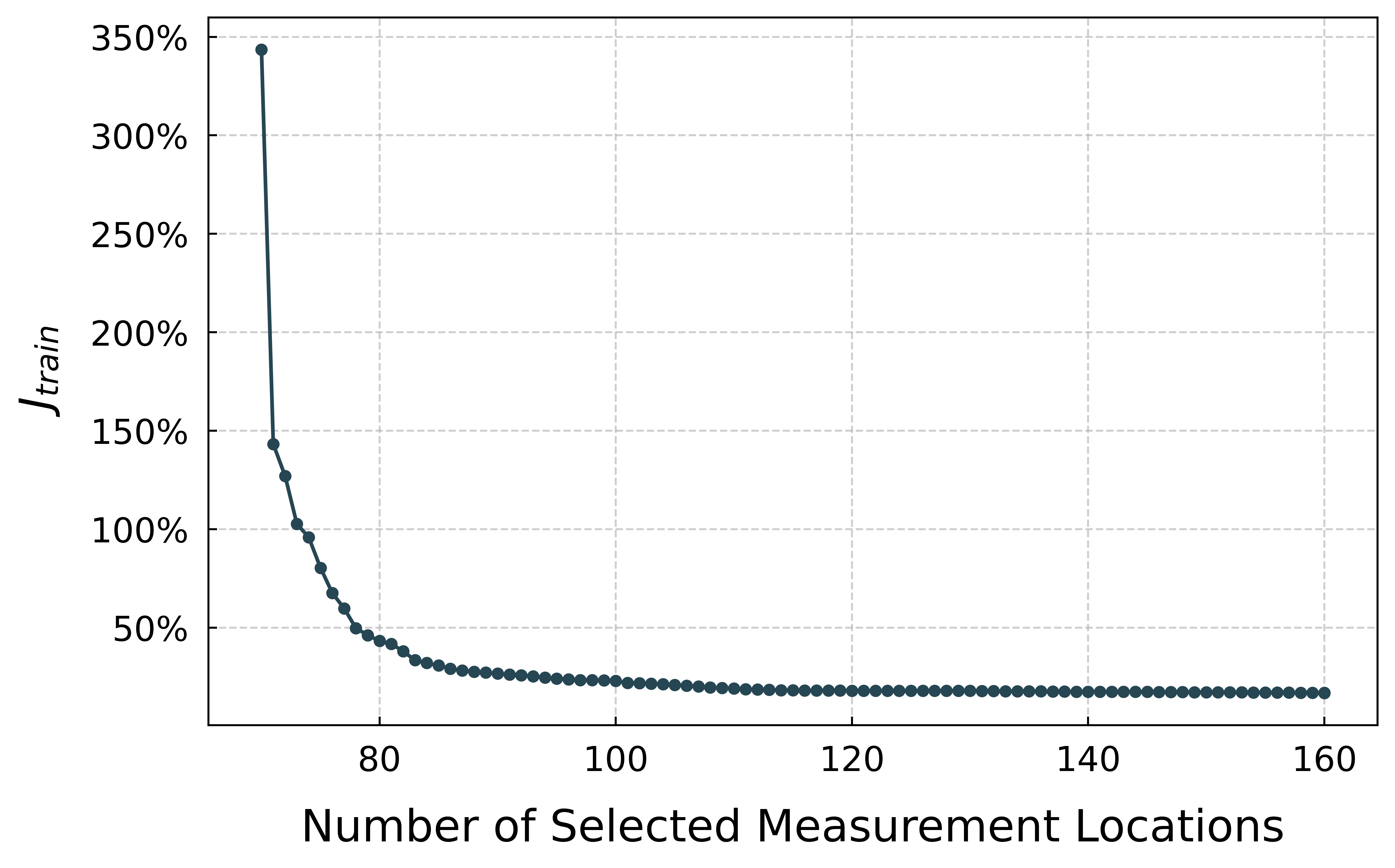}
\caption{Convergence of the training objective $J_{\mathrm{train}}$ with respect to the number of selected measurement locations using the proposed selection method.}
\label{fig:proposed_jtrain}
\end{figure}

In this study, the Absolute Relative Error (ARE), Mean Absolute Relative Error (MARE), and the Maximum Absolute Relative Error (MaxARE) are utilized as the evaluation metrics for assessing the efficacy of our framework. These metrics are defined as
\begin{align}
\mathrm{ARE}_{i,j}
&= \left|\frac{F_i^{\mathrm{pre}}(x_j)-F_i(x_j)}{F_i(x_j)}\right|
\tag{24}\label{eq:ARE}\\
\mathrm{MARE}
&= \frac{1}{n}\sum_{i=1}^{n}\left(\frac{1}{M}\sum_{j=1}^{M}\mathrm{ARE}_{i,j}\right)
\tag{25}\label{eq:MARE}\\
\mathrm{MaxARE}
&= \max_{1\le i\le n}\;\max_{1\le j\le M}\;\mathrm{ARE}_{i,j}
\tag{26}\label{eq:MaxARE}
\end{align}
in which $n$ is the number of cases, $F_i(x_j)$ is the OpenMC reference at node $x_j$ for case $i$, and $F_i^{\mathrm{pre}}(x_j)$ is the reconstructed value. The dataset is randomly split into a training subset (75\%) and a test subset (25\%), and 70 NPOD modes and the measurement location selection process are then obtained from the training subset. Therefore, the training and test dataset are almost in-distribution because they are sampled from the same scenario-generation process. Fig.~\ref{fig:proposed_jtrain} shows the convergence of the proposed selection method. The horizontal axis is the number of selected measurement locations, and the vertical axis is the training objective $J_{\mathrm{train}}$, corresponding to MaxARE on the training dataset. The objective decreases rapidly during the early selection stage and then gradually approaches a plateau. A total of 160 measurement locations is selected as the final measurement budget by considering the balance between MARE and MaxARE on the training dataset.

\begin{figure}[!t]
\centering
\includegraphics[width=\columnwidth]{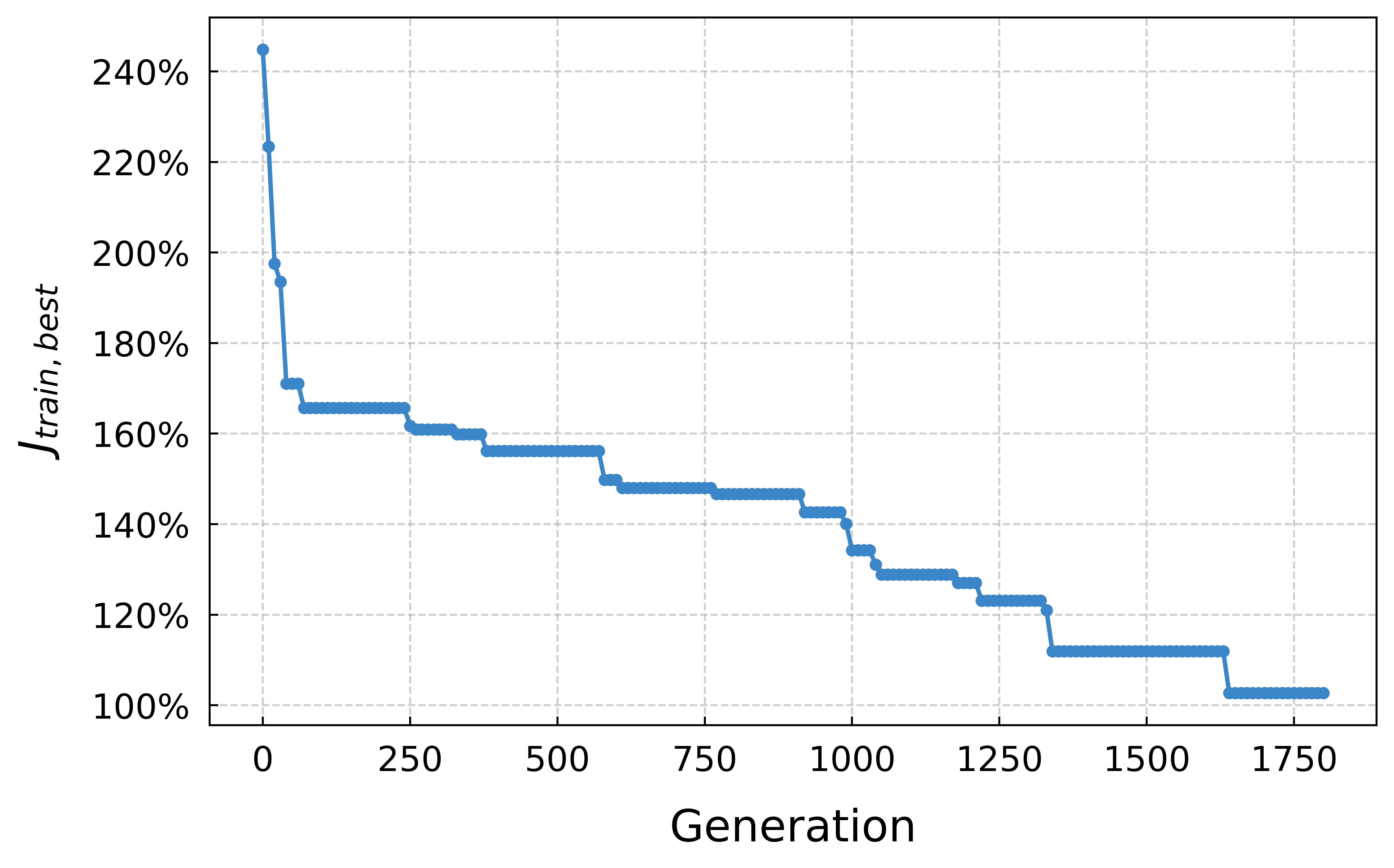}
\caption{Convergence of the best training objective $J_{\mathrm{train,best}}$ with respect to GA generation.}
\label{fig:ga_jtrain}
\end{figure}

Fig.~\ref{fig:ga_jtrain} shows the convergence of the GA baseline after hyperparameter tuning. To reduce the influence of arbitrary GA settings, a grid search is first performed over population size $\{100, 200, 400\}$, mutation rate $\{0.05, 0.10, 0.20\}$, elite fraction $\{0.05, 0.10\}$, and maximum generation number $\{100, 200, 300, 400,1000,1800\}$. These hyperparameter configurations are evaluated using the same NPOD basis with 160 measurement locations and the same training objective $J_{\mathrm{train}}$. The best-performing setting from this grid search is then used for the convergence run shown in Fig.~\ref{fig:ga_jtrain}, where the plotted quantity is $J_{\mathrm{train,best}}$, namely the best training objective in the population at each generation. The stepwise decrease is expected because $J_{\mathrm{train,best}}$ is updated only when a better measurement-location set is found during population evolution. Although the tuned GA continues to improve with more generations, the search remains challenging because it must identify a 160-location subset from 9948 admissible candidates. Table~\ref{tab:ga_minimax_comparison} summarizes the comparison between the tuned GA and the proposed selection method using optimizer-side objective evaluations and reconstruction accuracy. As shown in Table~\ref{tab:ga_minimax_comparison}, the proposed method uses 66.5\% fewer optimizer-side objective evaluations than the tuned GA baseline. More importantly, the proposed method substantially reduces MaxARE from 102.73\% to 20.51\%.
\begin{table}[!t]
\caption{Comparison between the tuned GA baseline and the proposed selection method.}
\label{tab:ga_minimax_comparison}
\centering
\small
\renewcommand{\arraystretch}{1.15}
\begin{tabular}{@{}lccc@{}}
\toprule
Metric                & GA baseline & Proposed & Reduction \\
\midrule
Objective evaluations & 648{,}400   & 217{,}461 & 66.5\% \\
MaxARE                & 102.73\%    & 20.51\%   & 80.0\% \\
\bottomrule
\end{tabular}
\end{table}

On the 1125 test cases, each containing predictions at 9948 spatial locations, the results generated by the proposed selection method show MARE of 0.28\% and a MaxARE of 20.51\%, indicating that the proposed method can effectively capture the complex gamma radiation distribution. These results indicate that the selected measurement locations and the NPOD-based inversion provide accurate reconstruction for the tested simulation distribution. Notably, when using a computing setup with 96 threads, 128GB RAM, and an AMD EPYC 7763 64-Core Processor to reconstruct 1125 test cases simultaneously, the process takes about 20 seconds in total. This equates to approximately 0.018 seconds per case, a speed that supports near-real-time reconstruction under this computing setup. In summary, the proposed NPOD-based framework demonstrates the capability for near real-time field reconstruction and a strong ability to capture nonlinear gamma distributions. Note that this verification case is simulation-based and uses noise-free selected measurements, so they do not explicitly represent detector noise during measurement acquisition. Therefore, an additional experiment-calibrated noisy case 1 test is performed. Specifically, the repeated measurements and background readings from our laboratory experiment described in Section~III-C are used to estimate a heteroscedastic detector-noise model, and the variance of a single count-rate reading is fitted as $\mathrm{Var}(y) = 0.0627y + 0.516$, where $y$ is the detector count rate. This noise model is then imposed only on the 160 selected measurement locations of each case 1 test field, while the train/test split, NPOD  basis, and selected measurement locations are kept fixed, and the noisy reconstructions are compared with the original clean OpenMC test fields. Table~\ref{tab:case1_noisy_bridge} summarizes the reconstruction results under clean and experiment-calibrated noisy measurement conditions. In this table, ``noisy selected values'' refer to selected measurements perturbed by the detector-noise model.
\begin{table}[!t]
\caption{Experiment-calibrated noisy Case 1 reconstruction results. WRMSE is reported in cps.}
\label{tab:case1_noisy_bridge}
\centering
\small
\renewcommand{\arraystretch}{1.15}
\begin{tabular}{@{}lcccc@{}}
\toprule
Input condition       & Solver & MARE   & MaxARE  & WRMSE \\
\midrule
Clean selected values & OLS    & 0.28\% & 20.51\% & N/A   \\
Noisy selected values & WLS    & 0.73\% & 37.68\% & 0.94  \\
Noisy selected values & OLS    & 1.02\% & 92.48\% & 2.63  \\
\bottomrule
\end{tabular}
\end{table}

The results show that laboratory-calibrated detector noise increases the reconstruction error in the large-scale Case 1 environment. However, the WLS reconstruction remains stable, with the MARE increasing from 0.28\% to 0.73\%. Compared with OLS, WLS substantially reduces the noise-induced instability, decreasing the MaxARE from 92.48\% to 37.68\% and the WRMSE from 2.63 to 0.94.  In the subsequent experimental study, a practical  measurement acquisition process with real detector data is used to further evaluate the robustness of the proposed method under noise and uncertainty.

\subsection{Case 2: 2-D Experiment Validation}

\begin{figure}[htbp]
  \centering
  \includegraphics[width=3in]{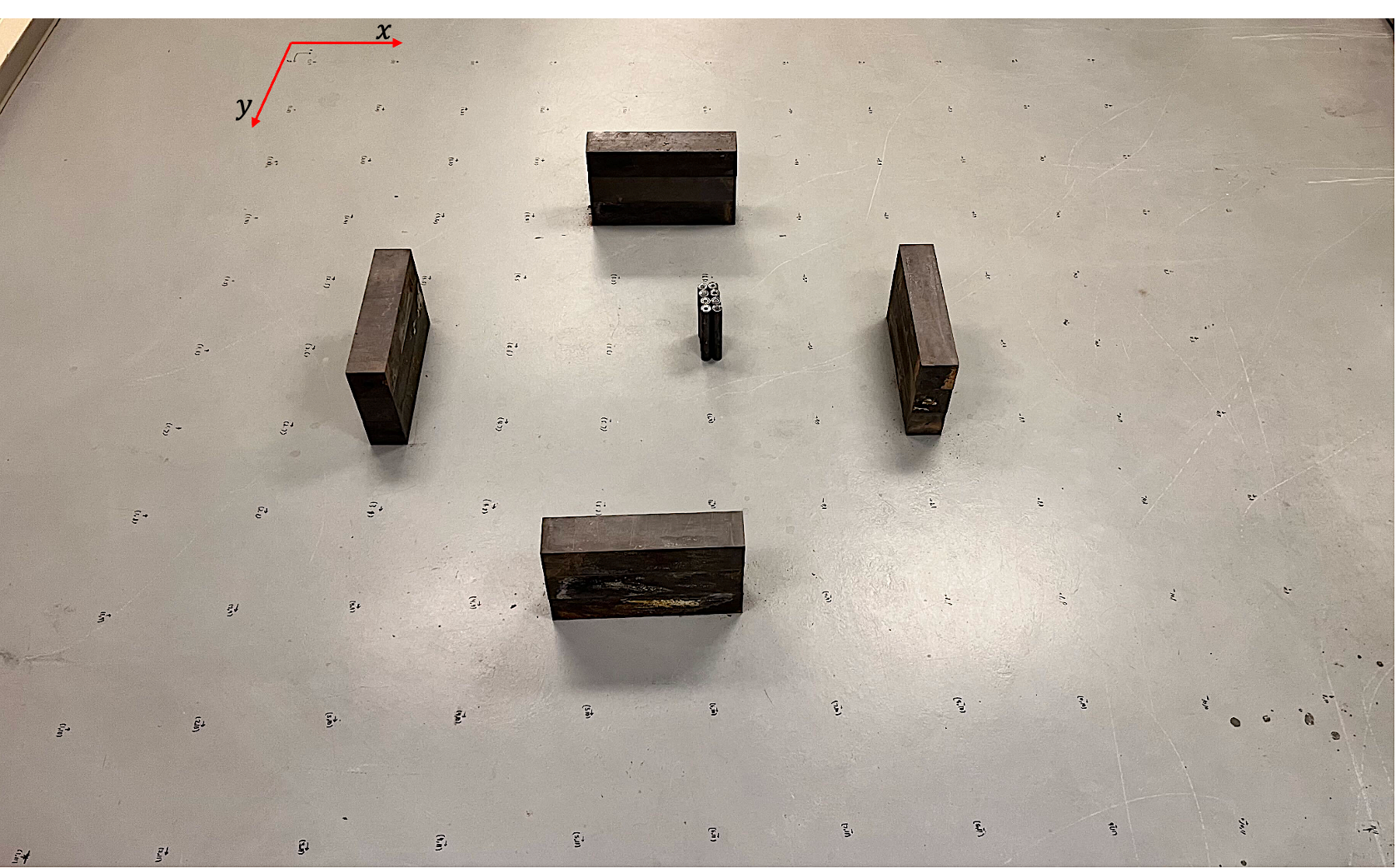}
  \caption{Experiment setup.}
  \label{fig:environment}
\end{figure}

\begin{figure}[htbp]
  \centering
  \includegraphics[width=3in]{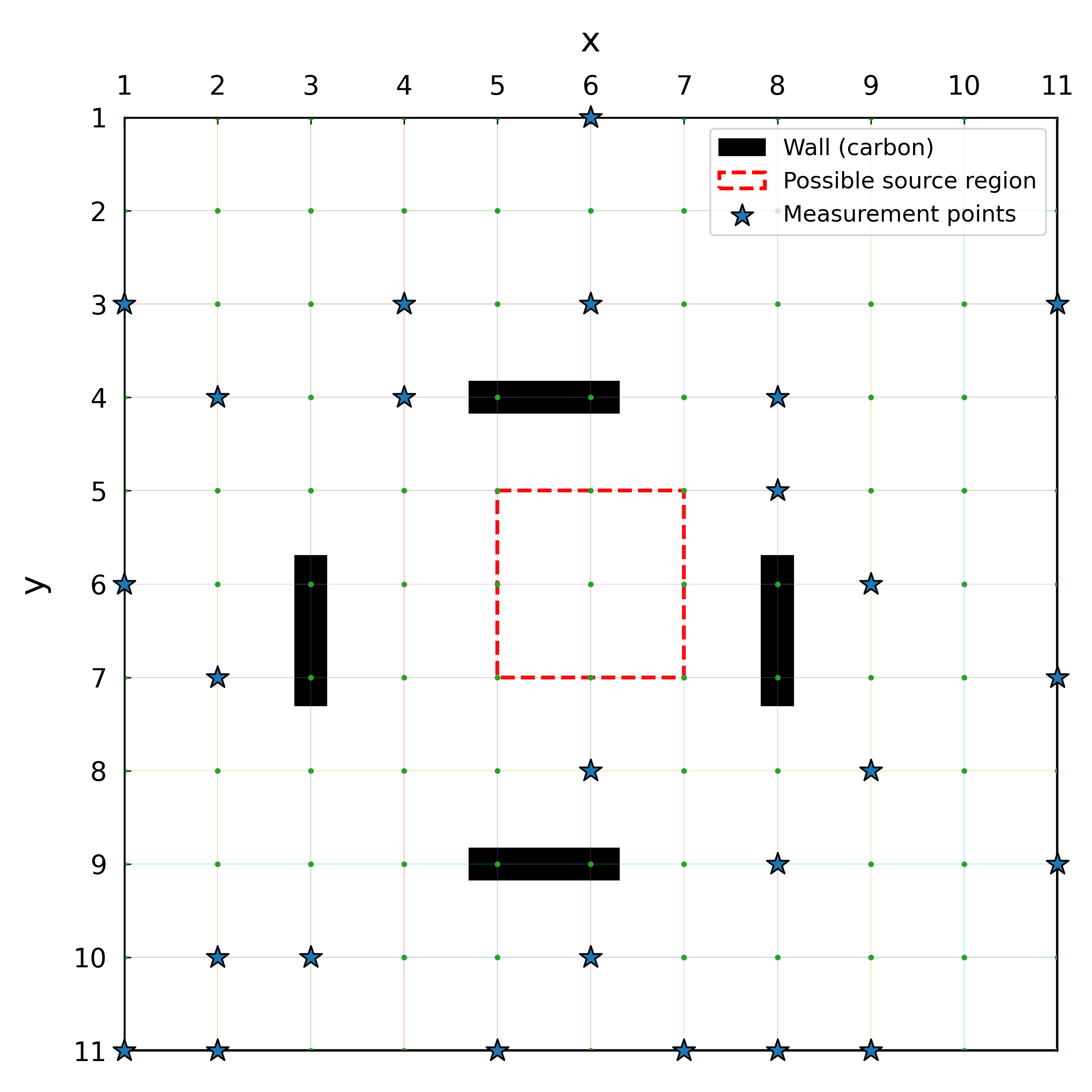}
  \caption{OpenMC top view of the environment, possible source region and final measurement locations.}
  \label{fig:top_view}
\end{figure}

\begin{figure}[htbp]
  \centering
  \includegraphics[width=3in]{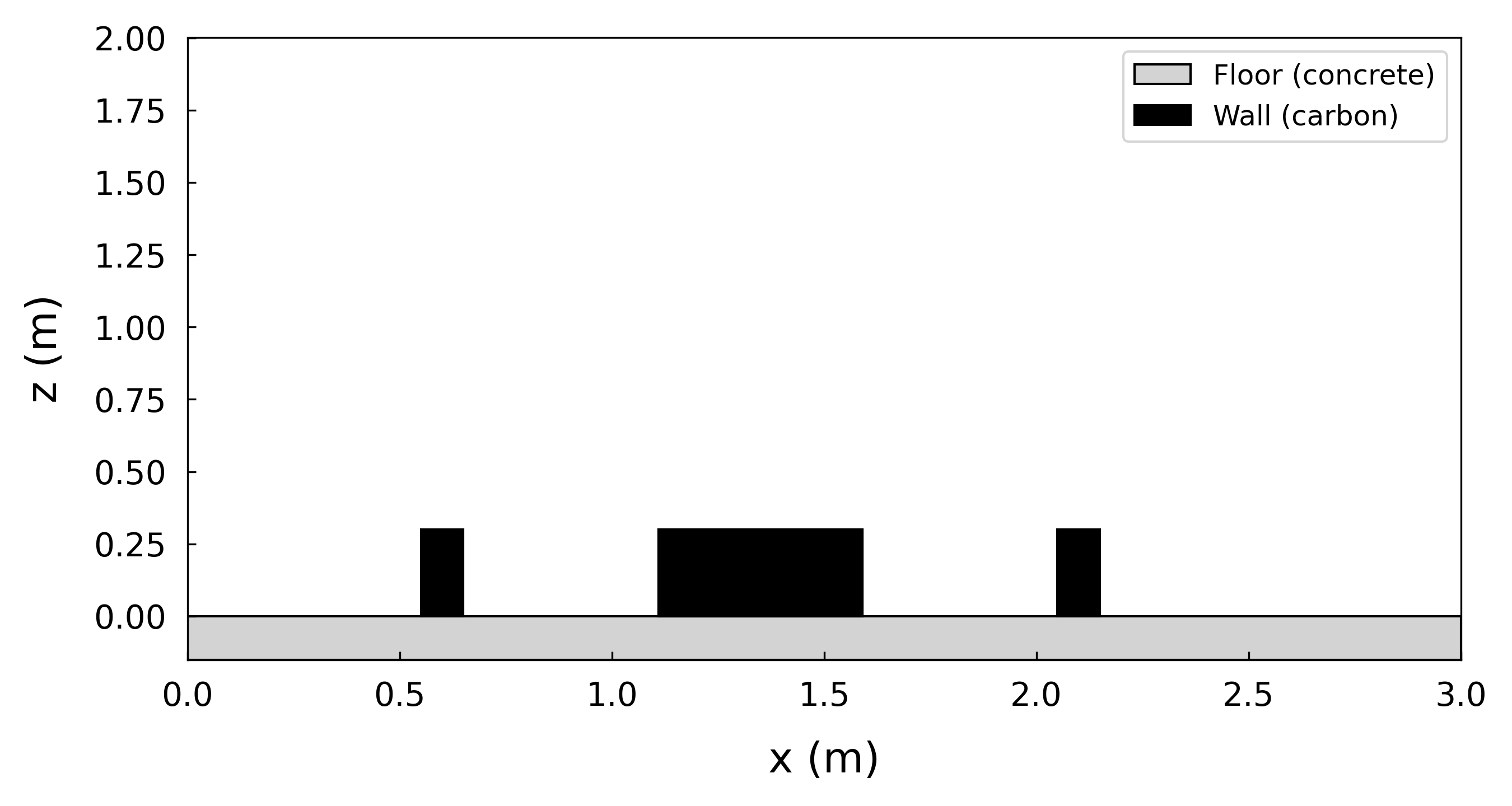}
  \caption{OpenMC front view of the environment.}
  \label{fig:front_view}
\end{figure}

The experiment is conducted in a square room of side $3.0\,\mathrm{m}$. The room is discretized on an $11\times11$ lattice with grid spacing $0.30\,\mathrm{m}$ ($i,j\in\{1,\ldots,11\}$). The physical environment is shown in Fig.~\ref{fig:environment}, and the top and front views in OpenMC simulations are shown in Figs.~\ref{fig:top_view} and \ref{fig:front_view}.

\begin{figure}[htbp]
  \centering
  \includegraphics[width=3in]{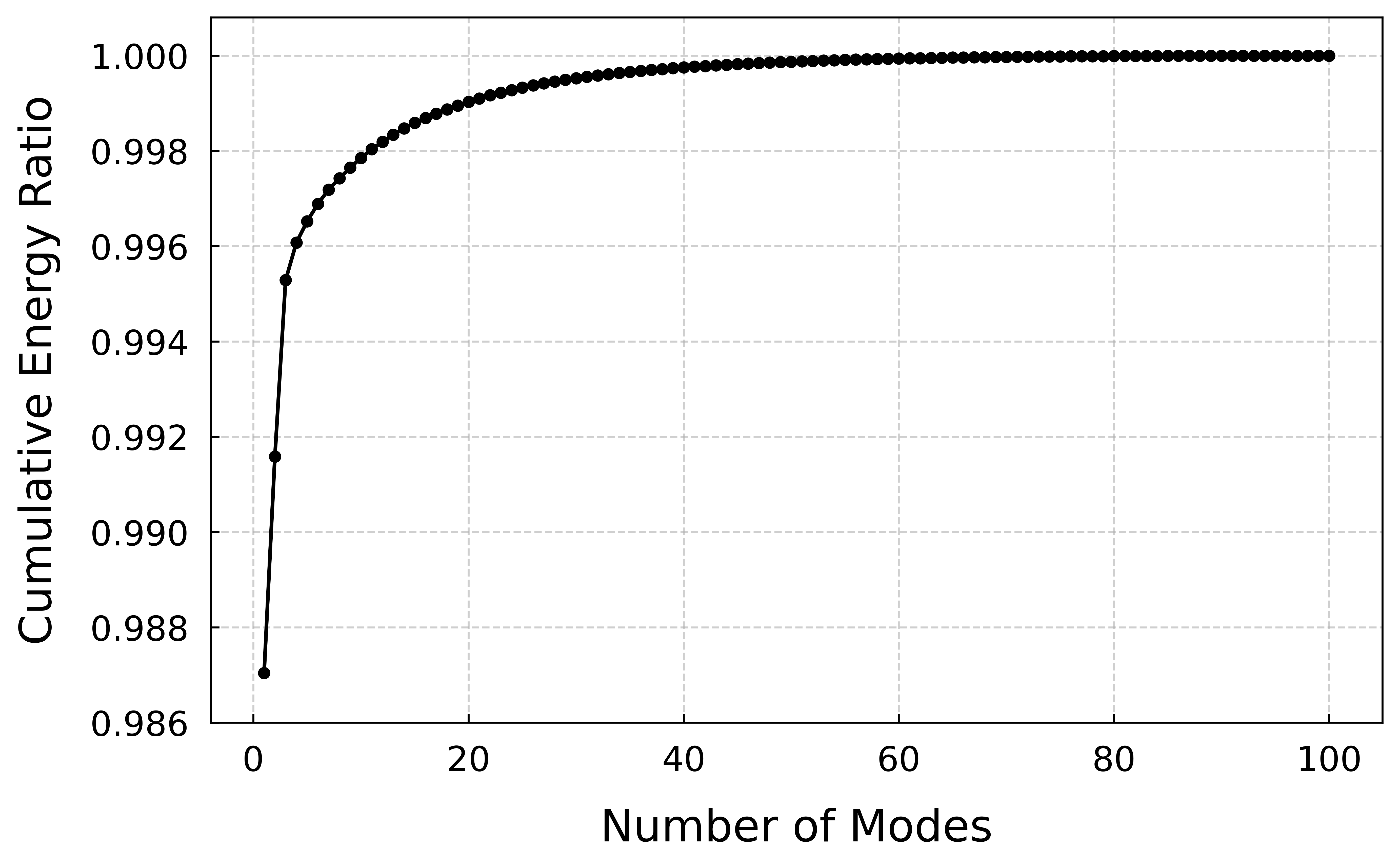}
  \caption{Cumulative energy of NPOD modes for the experiment.}
  \label{fig:pod_cum_ex}
\end{figure}

\begin{figure}[htbp]
  \centering
  \includegraphics[width=3in]{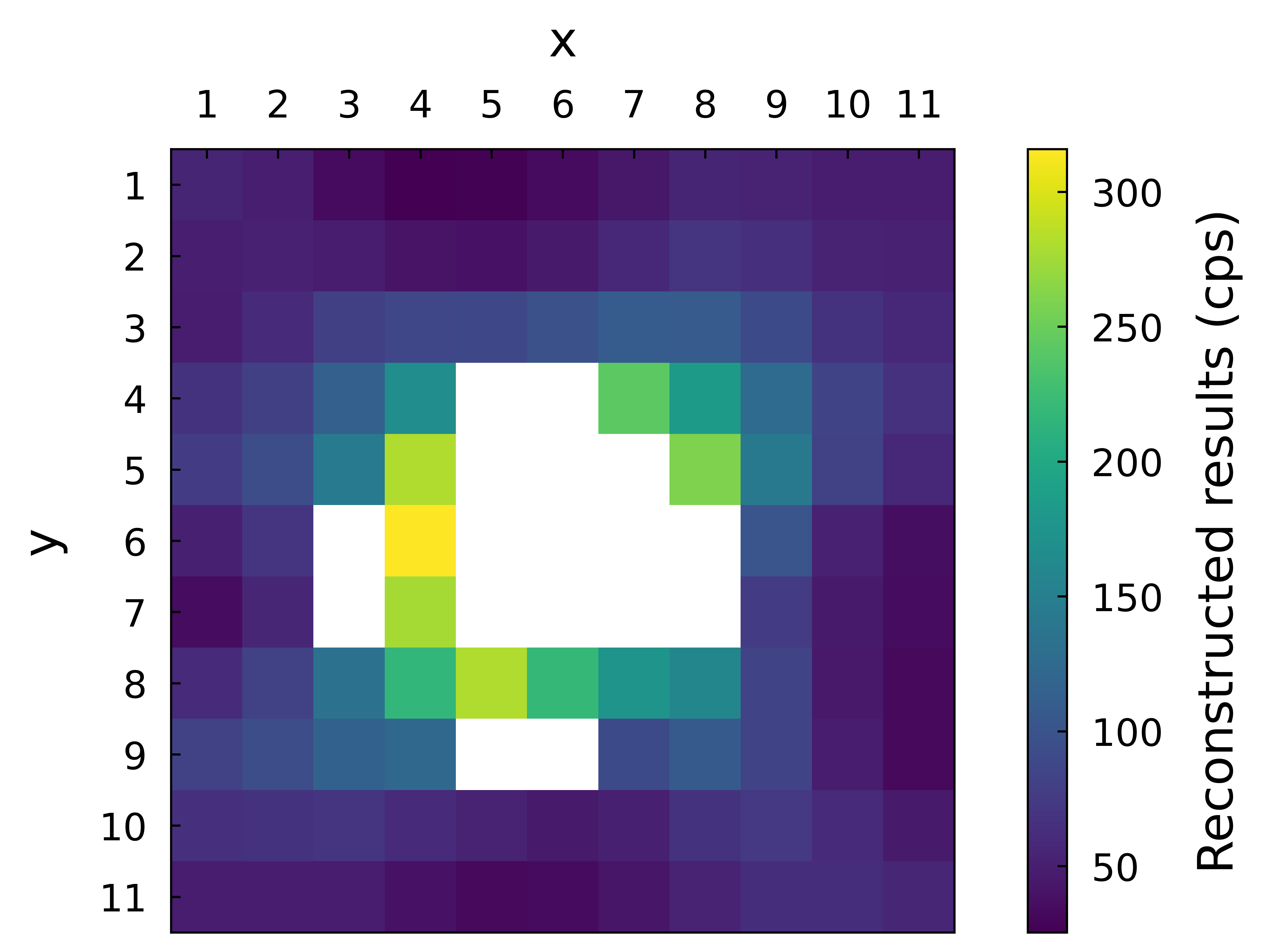}
  \caption{Reconstructed radiation field on the 2-D grid.}
  \label{fig:recon_heatmap}
\end{figure}

\begin{figure}[htbp]
  \centering
  \includegraphics[width=3in]{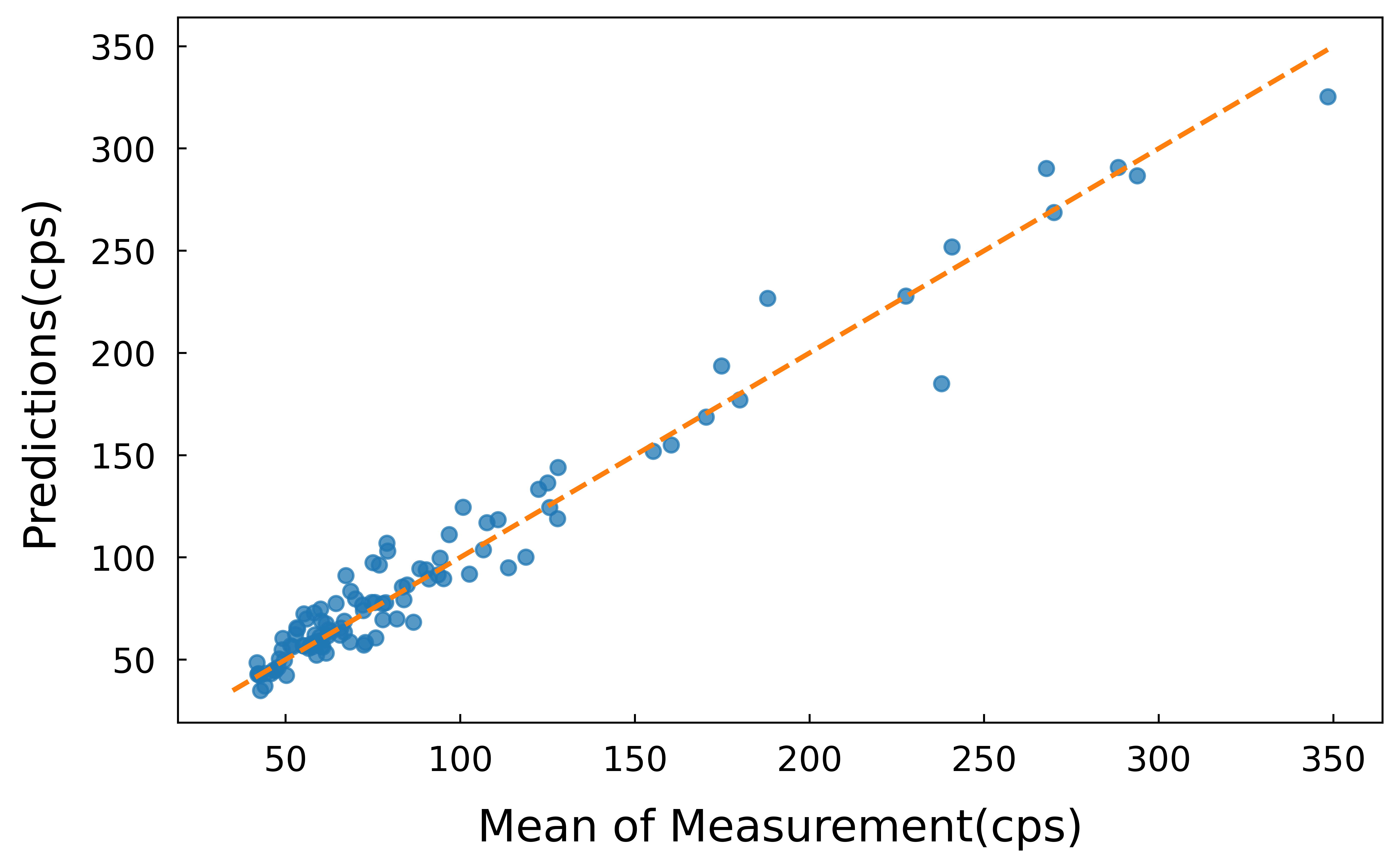}
  \caption{Measured versus predicted mean count rates with the identity line as reference.}
  \label{fig:meas_vs_pred_scatter}
\end{figure}

In the OpenMC model, a $0.15\,\mathrm{m}$ concrete floor lies on wet soil, and the upper boundary is at $z=2.0\,\mathrm{m}$ shown in Fig.~\ref{fig:front_view}. Four carbon wall segments of in-plane size $0.48\times0.10\,\mathrm{m}$ and height $0.30\,\mathrm{m}$ are placed along the lattice as shown in the top view. The central red dashed square in Fig.~\ref{fig:top_view} is reserved as a source region and is not used for measurements. Excluding walls and this region leaves $M=104$ admissible lattice points. To train NPOD for unknown source number, strength, and placement, $N=2000$ fixed-source cases are generated with OpenMC. Each case uses $n_s\in\{3,7,9,10,13\}$ identical cylindrical sources with diameter $2\,\mathrm{cm}$, height $24\,\mathrm{cm}$, $z\in[0,0.24]\,\mathrm{m}$, emitting $2540\,\gamma/\mathrm{s}$ per cylinder with a discrete spectrum $\{0.086,\,0.566,\,1.001\}\,\mathrm{MeV}$ and relative yields $\{0.70,\,0.25,\,0.05\}$. The physical experiment uses eight cylindrical sources located within the same source region. This eight-source configuration is not explicitly included in the source-count set used to construct the NPOD basis. Therefore, Case 2 provides a controlled out-of-distribution test with respect to source multiplicity. For each $n_s$, $400$ unique multi-source layouts are sampled randomly. Source centers are drawn without replacement from a $60\times60$ grid that tiles a $0.60\times0.60\,\mathrm{m}$ square and its lower-left and upper-right corners coincide with lattice nodes $(5,7)$ and $(7,5)$ in Fig.~\ref{fig:top_view}. Transport tallies are recorded on a $10\times10\times1$ mesh covering the measurement slice and are interpolated to the 104 admissible points by averaging the nearest cells. Flux is converted to counts per second (cps) using a fixed energy-weighted calibration derived from the detector datasheet. In the experiment, the detector is hand-carried at a fixed height to mimic a small mobile platform. The background without sources is first measured, and then five measurements are taken at each visited lattice point in the source-present environment. The cumulative energy distribution of NPOD modes for this case is shown in Fig.~\ref{fig:pod_cum_ex}. As shown in Fig.~\ref{fig:pod_cum_ex}, the number of modes is set to 20 due to $E_K >99.9$\%. The proposed location selection policy is used to choose the measurement location, and the final stops are marked in Fig.~\ref{fig:top_view}.
Fig.~\ref{fig:recon_heatmap} shows the reconstructed radiation field over the 2-D grid. The reconstructed field is smooth and physically consistent with the expected source and shielding layout, with larger gradients appearing near geometric constraints and exclusion zones. This suggests that the inferred NPOD coefficients capture the overall spatial pattern of the measured field. Table~\ref{tab:wrmse} reports the WRMSE for two settings and WLS achieves a substantially lower WRMSE. The reduction reflects two advantages: 1) noisy or low-count points are correctly down-weighted during fitting, and 2) the selected measurements are placed where information gain is high under the minimax criterion, which limits worst-case residuals that would otherwise dominate a weighted error.
Fig.~\ref{fig:meas_vs_pred_scatter} compares the measured mean count rates with the reconstructed values. The points cluster tightly around the identity line (dashed), indicating near-linear agreement across the full dynamic range. Deviations remain bounded even at high count rates, with only a small uniform offset at low counts. The measured mean count rate yields a MARE of 9.9\% and a MaxARE of 35.6\%.

\begin{table}[!t]
\caption{Comparison of WRMSE between uncertainty-aware WLS and uncertainty-ignorant OLS reconstructions. WRMSE is reported in cps.}
\label{tab:wrmse}
\centering
\small
\renewcommand{\arraystretch}{1.15}
\begin{tabular}{@{}lcc@{}}
\toprule
Solver & WLS  & OLS   \\
\midrule
WRMSE  & 3.61 & 13.30 \\
\bottomrule
\end{tabular}
\end{table}

\begin{figure}[htbp]
  \centering
  \includegraphics[width=3in]{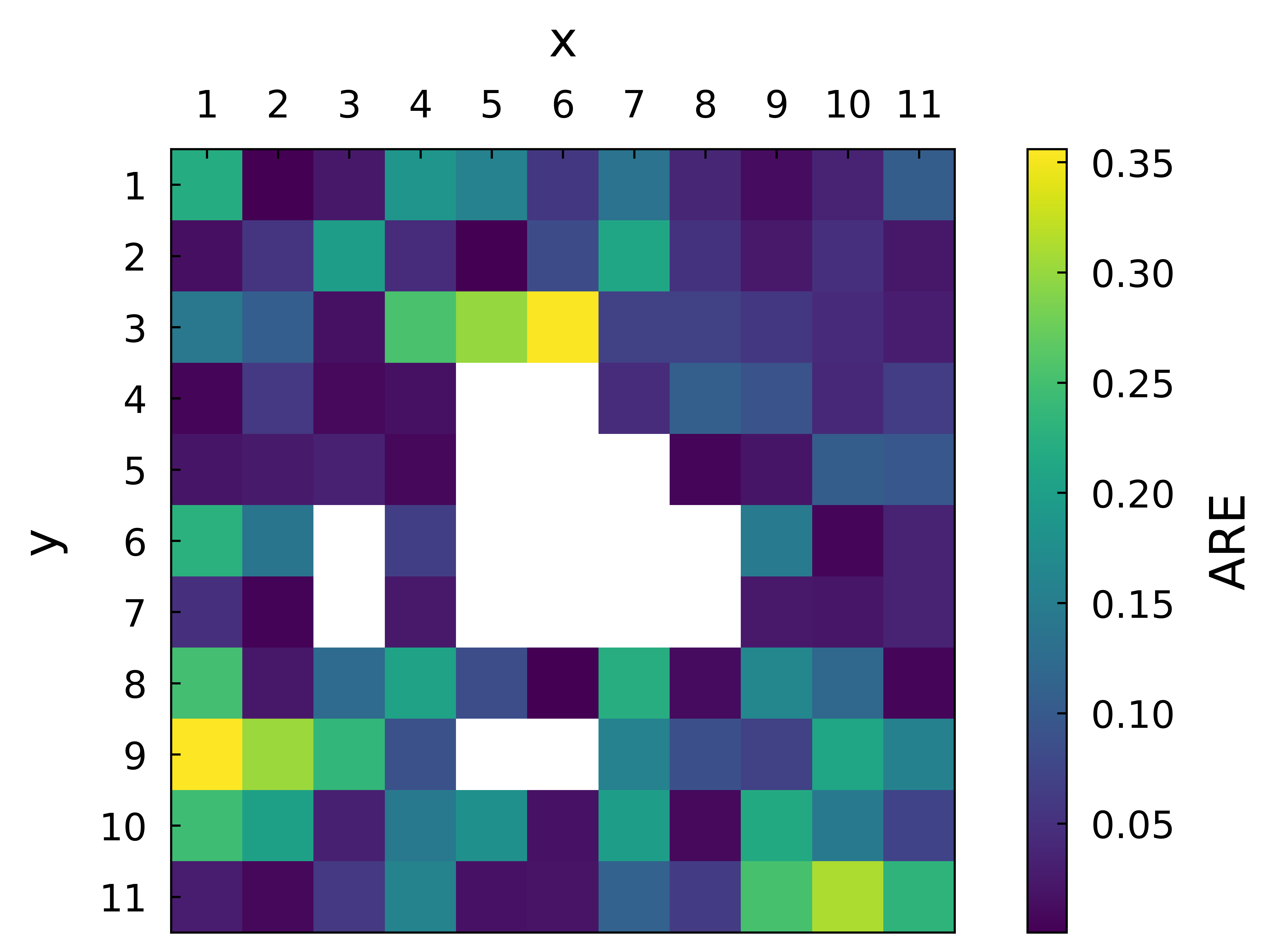}
  \caption{Reconstruction relative-error heatmap on measured core points.}
  \label{fig:recon_re_heatmap}
\end{figure}

Fig.~\ref{fig:recon_re_heatmap} displays the cell-wise ARE. Larger errors concentrate near rapid spatial transitions, around uninstrumented holes, and along boundaries where constraints are weaker. These areas either exhibit strong local gradients or lack nearby observations, so small absolute mismatches become large in relative terms. Regions near selected measurement points, especially those aligned with dominant NPOD modes, show uniformly low errors and smooth residuals.

\begin{figure}[htbp]
  \centering
  \includegraphics[width=3in]{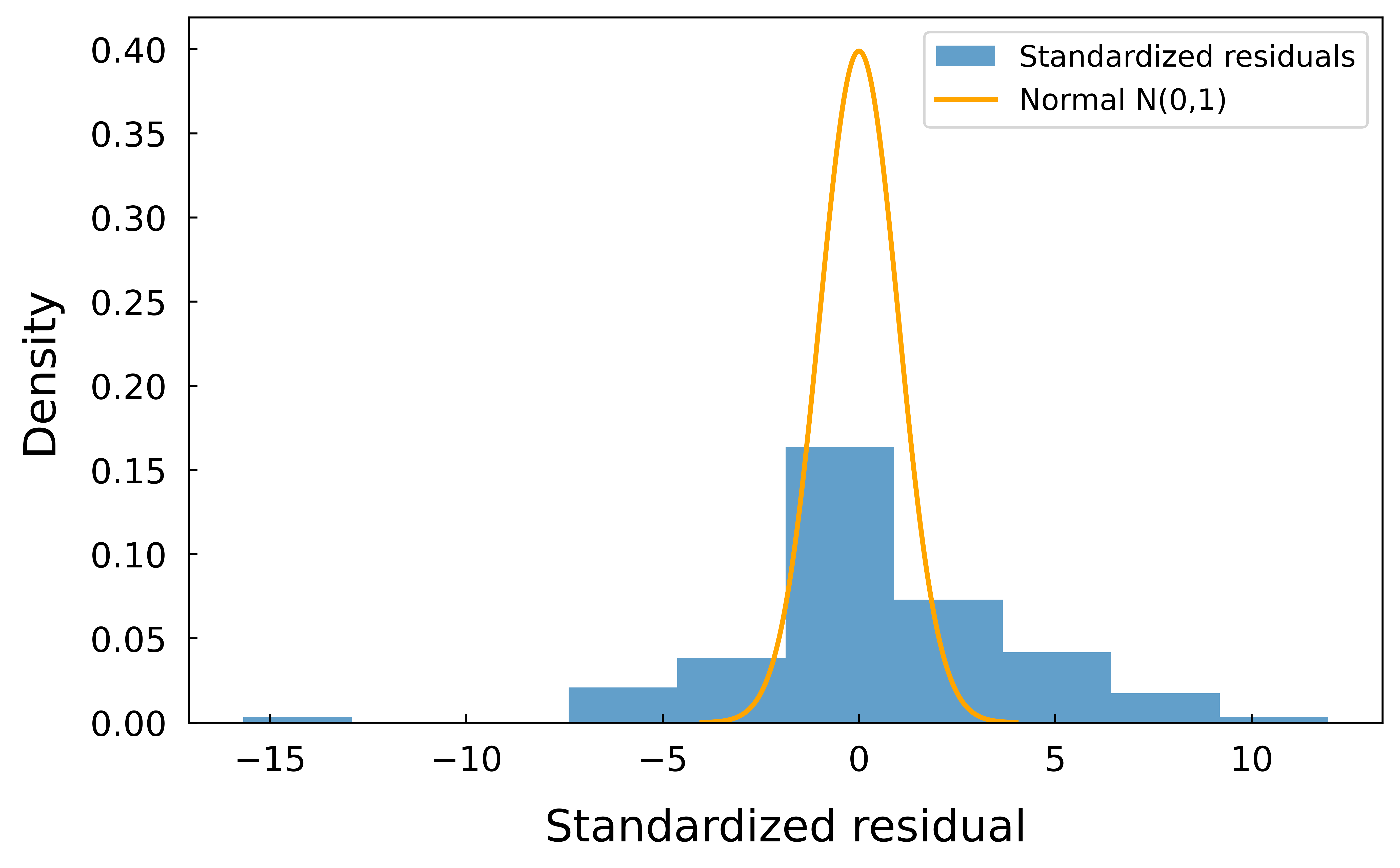}
  \caption{Histogram of standardized residuals in the core with the standard normal density as reference.}
  \label{fig:residual_hist}
\end{figure}

Fig.~\ref{fig:residual_hist} shows the histogram of standardized residuals. A standardized residual is the reconstruction error normalized by the predicted uncertainty at each spatial location. It can be interpreted as how many standard deviations the reconstruction deviates from the ground truth. If the uncertainty estimates are well calibrated and the residual behavior is approximately Gaussian, the standardized residuals should follow a standard normal distribution $\textit{N}(0,1)$. Therefore, the overlaid normal curve serves as a calibration reference. The distribution is approximately centered at zero, which suggests no systematic over- or under-prediction. Most residuals are close to the normal reference, indicating that the uncertainty calibration is reasonable for the majority of locations. The mildly heavier tails and a few outliers suggest that larger errors can occur at a small subset of points. These points are typically associated with high-gradient regions or domain boundaries, which is consistent with the relative-error map.

As an intermediate check, a nominal Case~2 OpenMC reconstruction is
performed using the fixed NPOD basis and the laboratory-matched OpenMC
field as the reference. It gives a MARE of 1.19\% and a MaxARE of
8.28\%, both higher than the in-distribution Case~1 results but below
the 9.9\% MARE observed in the physical experiment. Since the algorithm
and reconstruction procedure are unchanged between Case~1 and
the nominal Case~2 study, the increase from 0.28\% to 1.19\% is
primarily attributable to the out-of-distribution nature of the
source-count distribution in Case~2, where the laboratory configuration
of eight sources is not included in the training source-count set
$\{3,7,9,10,13\}$. To quantify the contribution to the error gap
between the nominal Case~2 simulation and the physical experiment, a
controlled OpenMC sensitivity study is performed. The NPOD basis and
the selected measurement locations are kept fixed, and each perturbed
scenario is averaged over 20 independently sampled perturbations. The
simulation reference and experimental reference rows in
Table~\ref{tab:case2_sensitivity} correspond to single cases.

\begin{table*}[!t]
\caption{Case~2 reconstruction-error sensitivity to controlled mismatch sources.}
\label{tab:case2_sensitivity}
\centering
\small
\renewcommand{\arraystretch}{1.25}
\begin{tabular}{@{}llp{0.40\textwidth}cc@{}}
\toprule
Type & Scenario & Main perturbation & MARE & MaxARE \\
\midrule
Simulation reference
  & Nominal Case~2 simulation         & None                                                                & 1.19\% & 8.28\%  \\
\cmidrule(l){1-5}
\multirow{4}{*}{\makecell[l]{Single-factor\\perturbations}}
  & Wall position                     & Carbon-wall centers shifted by $\Delta x,\Delta y \sim U(-3,3)$\,cm  & 3.21\% & 13.47\% \\
  & Material density                  & Carbon-wall and concrete densities varied by $\pm 5\%$               & 3.22\% & 13.51\% \\
  & Source position                   & Source centers shifted by $\pm 3$\,cm                                & 3.43\% & 16.04\% \\
  & Source strength                   & Individual source strengths varied by $\pm 10\%$                     & 3.15\% & 13.52\% \\
\cmidrule(l){1-5}
\multirow{3}{*}{\makecell[l]{Combined\\perturbations}}
  & Wall + material                   & Combined wall and material mismatch                                  & 3.31\% & 13.77\% \\
  & Wall + material + noise           & Combined mismatch with laboratory noise                              & 3.61\% & 16.66\% \\
  & All factors                       & All single-factor perturbations plus detector-chain uncertainty      & 5.55\% & 21.24\% \\
\cmidrule(l){1-5}
Experimental reference
  & Physical experiment               & Real setup                                                           & 9.90\% & 35.60\% \\
\bottomrule
\multicolumn{5}{@{}p{0.92\textwidth}@{}}{\vspace{2pt}\footnotesize
\textit{Note:} Detector-chain uncertainty includes $\pm 5\%$
calibration-scale uncertainty, $\pm 1.5\%$ detector-height response
variation, $\pm 1.7$\,cps background offset, and laboratory-calibrated
counting noise with $\mathrm{Var}(y)=0.0627\,y+0.516$.}
\end{tabular}
\end{table*}

Among the four single-factor perturbations, the MARE values
stay close to each other, ranging from 3.15\% to 3.43\%.
Source-position perturbation gives the largest single-factor MaxARE
of 16.04\%, while the other three single-factor MaxAREs lie between
13.47\% and 13.52\%.  Once factors are
combined, detector noise also contributes to MaxARE,
as seen in the wall + material + detector noise row at 16.66\%.
In addition, the combined-mismatch MARE of 5.55\% is sub-additive
relative to the sum of single-factor contributions. This is
physically expected because several mismatch effects are partially
correlated and can partially cancel within the WLS reconstruction.
For example, a wall-position shift can be partly absorbed by a
compensating density or source-strength adjustment in the NPOD
coefficients. This controlled study accounts for 4.36
percentage points of the 8.71-percentage-point gap from the nominal
Case~2 simulation at 1.19\% to the experiment at 9.9\%. The remaining approximately 4.35
percentage points reflect uncertainties not captured in the
controlled study, including exact source geometry and orientation,
detector angular response, nonuniform background, and hand-carried
detector placement variability. The 5.55\% value should therefore
be interpreted as a lower bound on the cumulative
simulation-to-experiment mismatch effect, not as a complete
decomposition.

\section{Conclusion}\label{sec:conclusion}

This paper introduces a framework for reconstructing gamma radiation fields that integrates NPOD with a minimax adaptive selection of measurement locations, offering both computational efficiency and reconstruction accuracy. Quantitative evaluations show consistently strong performance. In a large synthetic sweep involving 1125 test cases and 9948 spatial locations per case, the framework achieves a MARE of 0.28\% and a MaxARE of 20.51\% using only 160 measurement locations, while sustaining real-time reconstruction at $\sim$0.018\,s per case after the offline basis construction and measurement-location selection. In a 2-D laboratory experiment, the uncertainty-aware WLS formulation further reduces the WRMSE relative to a naive, uncertainty-ignorant variant and yields a near-identity measured-predicted scatter, indicating good calibration across the dynamic range. The higher experimental error is explained by a two-stage analysis which indicates that the difference is mainly associated with out-of-distribution source multiplicity and simulation-to-experiment mismatch.
The out-of-distribution source multiplicity mainly contributes to the gap from 0.28\% in Case 1 to 1.19\% in the nominal Case 2 reconstruction. The remaining gap from 1.19\% to 9.9\% is attributed to combined geometry, material, source, and measurement-chain mismatches. Controlled OpenMC perturbations capture approximately half of this remaining discrepancy, yielding a MARE of 5.55\%. Compared with a classical genetic algorithm baseline, the proposed adaptive placement requires fewer objective evaluations and attains a substantially lower MaxARE without relying on exhaustive search. It should be noted that NPOD is a snapshot-based reduced-order method, and its accuracy depends on the representativeness of the offline snapshot ensemble. If the deployment scenario falls outside the snapshot-design domain, reconstruction errors may increase and the offline snapshot ensemble should be updated accordingly. Several directions can further improve practicality. First, the present implementation reconstructs an energy-integrated gamma field rather than a fully energy-resolved field. Extending the snapshot construction and NPOD basis to multi-energy fields would improve realism for dose-centric applications. Second, the current measurement planning does not yet encode robot mobility, risk, or operational costs. Incorporating kinematic and safety constraints into the selection objective would make the framework more deployment-ready. Finally, extending the current static mapping setting to dynamic radiation fields will require faster data acquisition. A practical approach is to combine streaming fixed detectors with a mobile robot that samples a small set of informative locations, enabling near real-time tracking of evolving sources.

\appendix[Derivation of the Mean and Variance of an Exponentially Attenuated Intensity Field]
\label{app:attenuation}

\renewcommand{\theequation}{A\arabic{equation}}
\setcounter{equation}{0}

\subsection{Problem Statement and Notation}

\textbf{Geometry:} A one-dimensional slab of thickness $L>0$.

\textbf{Intensity law:} For a mono-energetic, collimated beam the transmitted intensity at depth $x$ is
\begin{equation}
I(x)=I_0 e^{-\mu x}, \qquad 0\le x\le L,
\label{eq:intensity}
\end{equation}
where $I_0>0$ is the unattenuated intensity at the incident surface and $\mu>0$ is the linear attenuation coefficient.

\textbf{Sampling model:} Assume the point of interest is selected uniformly at random along the slab,
\[
X\sim\mathcal{U}(0,L).
\]
Define the intensity random variable $I=I(X)=I_0 e^{-\mu X}$. Our goal is to derive
\[
\mathbb{E}[I],\quad
\mathbb{E}[I^{2}],\quad
\operatorname{Var}[I]=\mathbb{E}[I^{2}]-\bigl(\mathbb{E}[I]\bigr)^{2}.
\]

\subsection{First Moment (Mean Intensity)}
\begin{align}
\mathbb{E}[I]
  &=\int_{0}^{L} I_0 e^{-\mu x}\frac{1}{L}\,dx
   =\frac{I_0}{L}\int_{0}^{L} e^{-\mu x}\,dx \nonumber\\
  &=\frac{I_0}{L}
     \left[-\frac{1}{\mu}e^{-\mu x}\right]_{0}^{L}
   =\frac{I_0}{L\mu}\bigl(1-e^{-\mu L}\bigr).
\label{eq:mean}
\end{align}

\subsection{Second Raw Moment}
\begin{align}
\mathbb{E}[I^{2}]
  &=\frac{I_0^{2}}{L}\int_{0}^{L} e^{-2\mu x}\,dx
   =\frac{I_0^{2}}{L}
     \left[-\frac{1}{2\mu}e^{-2\mu x}\right]_{0}^{L} \nonumber\\
  &=\frac{I_0^{2}}{2\mu L}\bigl(1-e^{-2\mu L}\bigr).
\label{eq:second}
\end{align}

\subsection{Variance}
Substituting \eqref{eq:mean} and \eqref{eq:second},
\begin{align}
\operatorname{Var}[I]
  &=\frac{I_0^{2}}{2\mu L}\bigl(1-e^{-2\mu L}\bigr)
    -\left[\frac{I_0}{L\mu}\bigl(1-e^{-\mu L}\bigr)\right]^{2}\nonumber\\
  &=\frac{I_0^{2}}{2\mu L}\bigl(1-e^{-2\mu L}\bigr)
    -\frac{I_0^{2}}{(\mu L)^{2}}\bigl(1-e^{-\mu L}\bigr)^{2}.
\label{eq:variance}
\end{align}

\subsection{Consistency Checks}

Limit $L \to 0$: Using $e^{-\mu L}=1-\mu L+\frac{1}{2}\mu^2L^2+O(L^3)$,
both terms in (A4) approach $I_0^2$, and their difference tends to zero.
More specifically, a Taylor expansion gives
\[
\mathrm{Var}[I]\approx \frac{I_0^2\mu^2L^2}{12},
\]
so $\mathrm{Var}[I]\to 0$.
\textit{Limit $\mu L\gg1$:} With $e^{-\mu L}\approx0$ and $e^{-2\mu L}\approx0$,
\[
\operatorname{Var}[I]\simeq\frac{I_{0}^{2}}{2\mu L},
\]
i.e., the variance diminishes inversely with optical thickness.

\end{document}